\documentclass[%
 reprint,
 amsmath,amssymb,
prb,
]{revtex4-1}

\usepackage{graphicx}% Include figure files
\usepackage{dcolumn,subfigure}% Align table columns on decimal point
\usepackage{bm}% bold math
\usepackage{amsfonts}
\usepackage{amsmath}
\usepackage{amssymb}
\usepackage{amsthm}
\usepackage{array}
\usepackage{graphicx}
\usepackage{float}
\usepackage{booktabs,csquotes}

\usepackage{natbib,mathtools}
\usepackage{psfrag}
\usepackage{mathrsfs}

\newcommand{\rmi}{\mathrm{i}}

\newcommand{\rms}{\mathrm{s}}

\newcommand{\rmx}{\mathrm{x}}  
\newcommand{\rmy}{\mathrm{y}}  
\newcommand{\rmz}{\mathrm{z}}  
\newcommand{\rmA}{\mathrm{A}}

\newcommand{\rmO}{\mathrm{O}}

\newcommand{\rmS}{\mathrm{S}}

\newcommand{\bfD}{\mathbf{D}} 
\newcommand{\bfE}{\mathbf{E}}

\newcommand{\eff}{\mathrm{eff}}  
\newcommand{\art}{\mathrm{art}}  
\newcommand{\nm}{\mathrm{nm}}  

\newcommand{\xx}{\mathrm{xx}}  
\newcommand{\yy}{\mathrm{yy}}  
\newcommand{\zz}{\mathrm{zz}}  
\newcommand{\yz}{\mathrm{yz}}  
\newcommand{\xz}{\mathrm{xz}}  
\newcommand{\xy}{\mathrm{xy}}

\begin{document}

\preprint{APS/123-QED}

\title{Enhanced acousto-optic properties in layered media  }% Force line breaks with \\

\author{M. J. A. Smith}
\email{michael.j.smith@sydney.edu.au}
\affiliation{Centre for Ultrahigh bandwidth Devices for Optical Systems (CUDOS), Institute of Photonics and Optical Science (IPOS), School of Physics, The University of Sydney, NSW 2006, Australia}
 \author{C. Martijn de Sterke}
\affiliation{Centre for Ultrahigh bandwidth Devices for Optical Systems (CUDOS), Institute of Photonics and Optical Science (IPOS), School of Physics, The University of Sydney, NSW 2006, Australia}

\author{C. Wolff}
\author{M. Lapine}
\author{C. G. Poulton}
 \affiliation{School of Mathematical and Physical Sciences, University of Technology Sydney, NSW 2007, Australia }

\date{\today}

\begin{abstract} \noindent
We present a rigorous procedure for evaluating the photoelastic coefficients of a layered medium where the periodicity is smaller than the wavelengths of all optical and acoustic fields. Analytical expressions are given for the coefficients of a composite material comprising thin layers of optically isotropic materials. These coefficients include artificial contributions that are unique to structured media and arise from the optical and mechanical contrast between the constituents. Using numerical examples, we demonstrate that the acousto-optic properties of layered structures can be enhanced beyond those of the constituent materials. Furthermore, we show that the acousto-optic response can be tuned as desired.

\begin{description}
\item[PACS numbers]
77.65.Bn, 78.20.H-,81.05.Xj, 42.65.Es
\end{description}
\end{abstract}

 \maketitle

\section{Introduction} \label{sec:intro}
Since the first phenomonological descriptions of the photoelastic effect by Pockels \cite{pockels1890ueber,nelson1971theory,nelson1979electric}, acousto-optics has played a significant role in   optics and materials science. Acousto-optic effects are critical for radio-frequency modulators \cite{dixon1967photoelastic,nelson1979electric,eggleton2013inducing} and the photoelastic effect is frequently used  to determine stress distributions surrounding cracks and material defects \cite{newnham2004properties}. More recently, acousto-optics has found applications in modern nanophotonics: photoelasticity is the fundamental effect that underpins cavity optomechanics \cite{aspelmeyer2014cavity,djafari2016phoxonic} and Stimulated Brillouin Scattering (SBS), which is critical for a diverse range of devices such as ultra-narrow linewidth filters and high-resolution sensors \cite{kobyakov2010stimulated,powers2011fundamentals,eggleton2013inducing}. These devices, however rely on the existing, fixed, photoelastic response of the material platform, which in technologically-important cases can be   small \cite{biegelsen1974photoelastic,van2015interaction}. At the same time, SBS is problematic for optical fibre systems    \cite{peral1999degradation},  and so there is    considerable interest in both the suppression and the enhancement of photoelasticity, depending on the application.

 It is   well-known that composite materials, such as  layered media, can possess aggregate quantities that are markedly different from their constituents \cite{sipe1992nonlinear,milton2002theory}. Recent work \cite{smith2015electrostriction,smith2016control,smith2016stimulated} has shown that this principle applies to the acousto-optic properties of composites. In contrast to the intricate and exotic designs seen in the optical metamaterials community, layered materials are amongst the simplest structures to fabricate, yet  a complete picture of the acousto-optic properties of layered media   has not yet been reported. To the best of our knowledge, the only  other   literature  concerning the    photoelastic tensor of   layered media is by \citet{rouhani1986effective}, where analytical expressions for an orthorhombic composite  comprising orthorhombic layers were derived. However, nearly all of the expressions for the effective photoelastic coefficients are incomplete, as they do not include artificial photoelastic contributions (discussed below).

 It has been widely accepted  that acousto-optic  interactions in   uniform, non-piezoelectric dielectric media are   captured by  the photoelastic tensor $p_{ijkl}$ defined by 
 \begin{equation} \label{eq:conventionalpijkl}
 \Delta (\varepsilon^{-1})_{ij} = p_{ijkl} s_{kl},
 \end{equation}
 where $\Delta(\varepsilon^{-1})_{ij}$ denotes a   change in the    inverse  permittivity tensor, and $s_{kl}$ is the linear strain tensor for small displacements from equilibrium. In this definition,  the photoelastic tensor  is treated as symmetric with respect to the first and second index pairs, i.e. $p_{ijkl} = p_{(ij)(kl)}$. However, the  definition in \eqref{eq:conventionalpijkl}   is only sufficient to describe the interaction between electromagnetic and acoustic waves in   dielectrics possessing isotropic or cubic symmetry. This definition was sufficient  in early research on light-sound interactions,  since   the first solid materials examined were either of sufficiently high symmetry, or possessed low optical anisotropy \cite{eggleton2013inducing}. However, \citet{nelson1970new} established that this form of the photo-elastic response omitted the contributions of local rotations that   arise whenever shear waves propagate within the material; the effects of these local rotations on the permittivity tensor vanish for isotropic and cubic materials, but are non-zero for media that possess lower levels of structural symmetry such as tetragonal lattices\cite{nye1985physical}. This {\it roto-optic effect} can be strong compared to the symmetric photoelastic effect, and is directly related to the optical anisotropy of the material. The total photoelastic response of the material is given by  \cite{nelson1970new}
  \begin{align} \nonumber
 \Delta (\varepsilon^{-1})_{ij} &= P_{ijkl} \partial_l u_k \\ \label{eq:nonconventionalpijkl} 
 &=p_{ijkl} s_{kl} + r_{ijkl} r_{kl}
 \end{align}
 where $P_{ijkl}$ is the full photoelastic tensor, $\partial_l u_k = s_{kl} + r_{kl}$ denotes the gradient of the displacement vector, $p_{(ij)(kl)} $ $r_{(ij) \left[ kl\right]}$ are   the symmetric and antisymmetric  components of $P_{ijkl}$, respectively,  and $r_{kl}$ is the infinitesimal rotation tensor (where   round and square bracket notations represent symmetric and antisymmetric index pairs,  following \citet{nelson1970new}, and we now omit   bracket notation on   index pairs for convenience). The definition   \eqref{eq:nonconventionalpijkl}    captures the potentially  large influence that the antisymmetric component of the photoelastic tensor $P_{ijkl}$ (otherwise known as the roto-optic tensor) can have on the scattering of light by an acoustic shear wave \cite{nelson1970new}.

 The analytic form of the roto-optic tensor in   uniform materials was given in  \cite{nelson1970new,sapriel1979acousto,nelson1979electric} where the tensor coefficients were found to be  directly linked to the optical anisotropy of the medium (for materials that do not possess monoclinic or triclinic symmetry).  Subsequently, it is important to consider the effects of both strains and rotations    when studying   acousto-optic interactions in optically anisotropic materials. Although a wide selection of natural uniform materials exhibit strong optical anisotropy (such as calcite \cite{nelson1972brillouin}), it is also possible to achieve selective control over the optical birefringence of a medium by   constructing  composite materials\cite{milton2002theory}.

In recent years, it has also been established that  the photoelastic properties of  structured materials exhibit a unique effect known as artificial photoelasticity.  This effect was first recognized in  composite materials comprising   cubic arrays of spheres suspended in an otherwise uniform material by \citet{smith2015electrostriction,smith2016control,smith2016stimulated}.    Artificial photoelasticity can be physically understood as follows;  under a finite strain, the different mechanical responses of the  constituent materials  alters the filling fraction, and in turn, contributes to changes in the   permittivity of the composite.   Such artificial contributions have been shown to play a significant role in the photoelastic   properties of composites \cite{smith2016control} and cannot be omitted, even for high symmetry structured materials.   

The two main contributions to an acousto-optic interaction are photoelasticity, describing changes in permittivity induced from bulk strains, and {\it moving boundary effects}, describing permittivity changes due to boundary strains (e.g. the boundary between a waveguide or a cavity and the surrounding air) \cite{nelson1979electric,wolff2015stimulated}.  There is an extensive literature examining  interface motion (moving boundary) contributions in acousto-optics for  layered media \cite{rouhani1991localised,matsuda2002reflection,schneider2013defect}, periodic structures\cite{chan2012optimized,rolland2012acousto}, and general structures \cite{johnson2002perturbation,wolff2015stimulated}, for example. However, the precise relationship between   the moving boundary effect  and artificial photoelasticity is presently unclear. Both effects relate to interface motions and   both require a permittivity contrast in order to feature in an acousto-optic  interaction. However, if the stiffness tensors of all layers are identical  $C_{ijkl} = C_{ijkl}^\prime$  then artificial photoelasticity is zero, whereas moving boundary contributions are not necessarily vanishing \cite{rouhani1991localised,wolff2015stimulated}. 

In place of photoelasticity and the moving boundary effect, it is also possible to describe   acousto-optic interactions in terms of  electrostriction, which describes bulk stresses induced by an electromagnetic field, and radiation pressure, describing  boundary stresses across dielectric interfaces  \cite{nelson1979electric}.  Analytical expressions for   the electrostrictive response of structured materials, under the approximation that the shear contribution is negligible, were given in \citet{smith2015electrostriction}, and   a rigorous numerical investigation followed soon after in \citet{smith2016control,smith2016stimulated}. In all instances, the electrostrictive properties of the composite were observed to be enhanced above and  beyond the intrinsic electrostrictive properties of the constituents, indicating that strong effects may also be observed in structured materials with reduced symmetry, such as layered media.

\begin{figure}[t]
\centering
 \label{fig:schematic}
\includegraphics[scale=0.49]{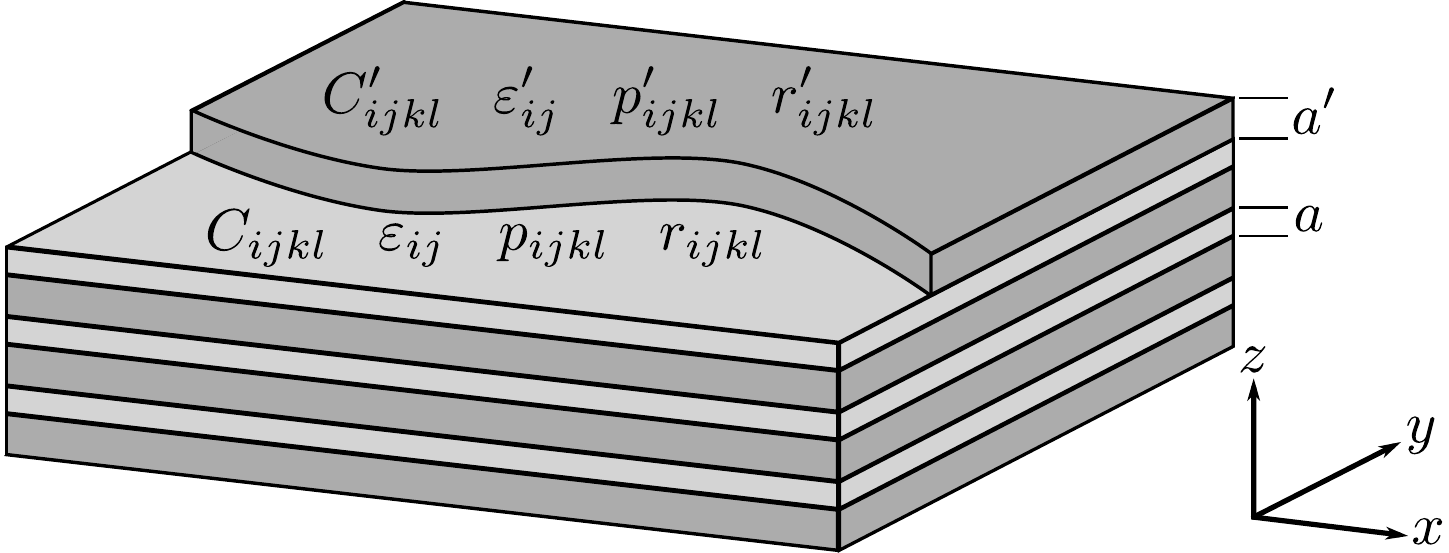} 

\caption{ Schematic of   layered material investigated (infinitely extending in the $x-y$ plane) with periodicity along $z$-axis and constituent parameters labelled. }
  \end{figure}

 In this paper, we derive the photoelastic coefficients of a layered medium, as shown in Fig. \ref{fig:schematic}, giving  the   artificial contribution to the symmetric photoelastic tensor explicitly, in addition to an explicit representation for the roto-optic tensor. These expressions are obtained from the closed-form expressions for the effective permittivity and    stiffness tensors,  where we do not consider frequency dependence in the materials properties \cite{nelson1977dispersive}.  The procedure we outline for the effective permittivity tensor is a generalisation of that presented in \citet{bergman1978dielectric}, which was   extended to   the effective stiffness tensor   by   \citet{smith2016stimulated}, and is analogous to   the approach   by \citet{grimsditch1985effective}.   We demonstrate photoelastic coefficients with values above and beyond that of either constituent material, strong roto-optic coefficients, and  non-negligible contributions from artificial photoelasticity, for a silica-silicon, and a silica-chalcogenide glass medium.

 The outline of this paper is as follows. In Section \ref{sec:effperm} we present the procedure for calculating the effective permittivity tensor $\varepsilon_{ij}^\eff$. In Section \ref{sec:effCijklsec} we consider the analogous procedure for the effective stiffness tensor $C_{ijkl}^\eff$. In Section \ref{sec:pijkleff} we determine the symmetric photoelastic coefficients $p_{ijkl}^\eff$, and in Section \ref{sec:pijklasymmeff} the antisymmetric photoelastic coefficients $r_{ijkl}^\eff$. This is followed by a  numerical study of  layered materials in Section \ref{sec:numerexamp} before concluding remarks in Section \ref{sec:conclrem}.

\section{Effective material parameters} \label{sec:effall} \noindent
In this section,  we   outline a compact procedure for calculating effective materials tensors, starting with the effective permittivity tensor\cite{smith2016stimulated,bergman1978dielectric},   and    the effective stiffness tensor\cite{smith2016stimulated}. In this work, the   layered medium is  constructed as  a one-dimensional stack of optically isotropic    dielectric slabs, with periodicity in $z$, that forms a medium with  tetragonal ($4/mmm$) symmetry \cite{nye1985physical} as shown in Fig. \ref{fig:schematic}. Results for $p_{ijkl}$  are presented explicitly for this case, although the procedure   is readily generalisable to consider layered materials made with optically anisotropic constituents.

The effective medium procedure outlined here essentially replaces the layered material with a hypothetical  effective material    exhibiting  the same boundary information on the edges of the unit cell, and possessing the  same   energy    as the layered material per  unit cell.  It is assumed that acousto-optic interactions are the only nonlinear effect that the effective medium exhibits.   In the derivation that follows, we use the convention of unprimed notation for the first layer in the unit cell, primed ($\prime$) notation for the second layer and `$\eff$' for the effective medium. It is assumed that the thicknesses of the two layers, $a$ and $a^\prime$, are  small relative to the   wavelength of all electromagnetic and acoustic fields (see Fig. \ref{fig:schematic}).  In other words, we examine the intrinsic bulk properties of the material in both the optical and acoustic long-wavelength regime. 

 It is also assumed that the optical and acoustic  contrast between layers does not induce a perturbation to the magnetic field, i.e., $\mu_{ij} = \mu_{ij}^\prime = \mu_{ij}^\eff = \delta_{ij}$, where $\delta_{ij}$ denotes the Kronecker delta and $\mu_{ij}$ the relative permeability.

\subsection{Effective permittivity tensor} \label{sec:effperm} \noindent
We begin by computing   the effective permittivity tensor for a layered medium and    impose conventional electromagnetic boundary conditions across the layers; continuity of the tangential $\bfE$ field and normal $\bfD$ field for our layered medium requires that
\begin{equation} \label{eq:maxwellcont}
E_\rmx = E_\rmx^\prime, \; E_\rmy = E_\rmy^\prime,  \; D_\rmz = D_\rmz^\prime,
\end{equation}
where we further impose  that the effective medium must   take the same static field values  at all boundaries
\begin{equation}
\label{eq:EMBC}
E_\rmx^\eff = E_\rmx = E_\rmx^\prime, \; E_\rmy^\eff = E_\rmy = E_\rmy^\prime, \;   D_\rmz^\eff =   D_\rmz = D_\rmz^\prime.
\end{equation}
We then  require that the effective energy density  \cite{jackson1962classical}
\begin{equation}
\mathcal{U}^\eff = \frac{1}{2}    E_i^\eff D_i^\eff,
\end{equation}
is equivalent to the total energy density   over the unit cell
\begin{equation}
\mathcal{U} = \frac{1}{2} \bigg( f E_i D_i +  (1-f) D_i^\prime E_i^\prime \bigg),
\end{equation}
where $f = a/(a+a^\prime)$ is the volume filling fraction, which  gives rise to    
\begin{subequations}
\label{eq:ENDENCall}
\begin{align}
\label{eq:ENDENC1}
D_\rmx^\eff &= f D_\rmx + (1-f) D_\rmx^\prime , \\
\label{eq:ENDENC2}
D_\rmy^\eff &= f D_\rmy + (1-f) D_\rmy^\prime , \\
\label{eq:ENDENC3}
E_\rmz^\eff &= f E_\rmz + (1-f) E_\rmz^\prime. 
\end{align}
\end{subequations}
Using   \eqref{eq:EMBC} and \eqref{eq:ENDENCall}   with the constitutive relations 
\begin{equation}
 D_i^\eff = \varepsilon_0 \varepsilon_{ij}^\eff E_j^\eff, \; D_i = \varepsilon_0 \varepsilon_{ij} E_j, \; D_i^\prime = \varepsilon_0 \varepsilon_{ij}^\prime E_j^\prime,
\end{equation}
where $ {\varepsilon}_{ij}$ denotes the permittivity tensor and $\varepsilon_0$ the vacuum permittivity, it follows almost immediately that 
\begin{subequations}
\label{eq:effeps}
\begin{align}
\label{eq:effepsxx}
\varepsilon_{\xx}^\eff   &= f  \varepsilon_{\xx}    + (1-f)    \varepsilon_{\xx}^\prime  
- \frac{f(1-f)(\varepsilon_\xz - \varepsilon_\xz^\prime)^2}{ f \varepsilon_\zz^\prime +  (1-f) \varepsilon_\zz}, \\
\label{eq:effepsyy}
\varepsilon_{\yy}^\eff   &= f  \varepsilon_{\yy}    + (1-f)    \varepsilon_{\yy}^\prime  
- \frac{f(1-f)(\varepsilon_\yz - \varepsilon_\yz^\prime)^2}{ f \varepsilon_\zz^\prime +  (1-f) \varepsilon_\zz}, \\
\label{eq:effepszz}
\frac{1}{\varepsilon_\zz^\eff} &= \frac{f}{\varepsilon_\zz} +  \frac{(1-f)}{\varepsilon_\zz^\prime}, \\
\label{eq:effepsyz}
\varepsilon_\yz^\eff &= \frac{ f \varepsilon_\yz \varepsilon_\zz^\prime   + (1-f) \varepsilon_\yz^\prime \varepsilon_\zz  }{ f \varepsilon_\zz^\prime + (1-f) \varepsilon_\zz}, \\
\label{eq:effepsxz}
\varepsilon_\xz^\eff &= \frac{ f \varepsilon_\xz \varepsilon_\zz^\prime   + (1-f) \varepsilon_\xz^\prime \varepsilon_\zz  }{ f \varepsilon_\zz^\prime + (1-f) \varepsilon_\zz}, 
\end{align}
\begin{multline}
\label{eq:effepsxy}
 \varepsilon_{\xy}^\eff     = f      \varepsilon_{\xy}         + (1-f)    \varepsilon_{\xy}^\prime    \\ 
  - \frac{f(1-f)(\varepsilon_\xz - \varepsilon_\xz^\prime)(\varepsilon_\yz - \varepsilon_\yz^\prime)}{ f \varepsilon_\zz^\prime +  (1-f) \varepsilon_\zz}.
\end{multline}
\end{subequations}
The expressions for $\varepsilon_{ij}^\eff$ above are equivalent to those   presented in \citet{rouhani1986effective}, and are valid for layered materials comprising fully anisotropic layers.   Despite the equivalence of certain $\varepsilon_{ij}^\eff$ coefficients in a tetragonal material (i.e.,     $\varepsilon_\xx^\eff = \varepsilon_\yy^\eff$ and $\varepsilon_{\yz}^\eff =  \varepsilon_{\xz}^\eff = \varepsilon_{\xy}^\eff = 0$),  all permittivity coefficients  are required to determine   the photoelastic coefficients for the composite    in Section \ref{sec:pijkleff}. For reference,   we represent elements of an inverse tensor   by $(\varepsilon^{-1})_{ij}$ and reciprocal values by   $1/\varepsilon_{ij}$.

\subsection{Effective stiffness tensor} \label{sec:effCijklsec}\noindent
We now obtain closed-form expressions for the stiffness tensor of a layered material, and begin by  imposing conventional     acoustic boundary conditions \cite{auld1973acoustic}; continuity of transverse velocity (or transverse displacement for time-harmonic fields in the long-wavelength limit) and continuity of the normal component of the stress field, which requires that 
 \begin{align}
\label{eq:contdisptrans}
 u_\rmx  =  u_\rmx^\prime, \quad   u_\rmy =  u_\rmy^\prime, 
\end{align}
in addition to
\begin{equation}
\sigma_\xz  = \sigma_\xz^\prime, \quad \sigma_\yz = \sigma_\yz^\prime, \quad  \sigma_\zz = \sigma_\zz^\prime,
\end{equation}
respectively.   We then impose   that the effective displacement and effective stress fields possess the same  static  values at the boundary  as per the conditions above, for example, $ u_\rmx^\eff =  u_\rmx =  u_\rmx^\prime$
and $\sigma_\xz^\eff = \sigma_\xz  = \sigma_\xz^\prime$. In analogy to Section \ref{sec:effperm}, we require that  the strain energy density for the effective medium
\begin{equation}
\label{eq:strainendeneffeq}
\mathcal{U}_\rms^\eff = \frac{1}{2}   \sigma_{ij}^\eff s_{ij}^\eff,
\end{equation}
is equivalent to the total strain energy density
\begin{equation}
\mathcal{U}_\rms = \frac{1}{2}  \bigg( f \sigma_{ij} s_{ij} + (1-f) \sigma_{ij}^\prime s_{ij}^\prime \bigg),
\end{equation}
where $s_{ij} = \tfrac{1}{2}(\partial_i u_j + \partial_j u_i)$. This   is satisfied provided   
\begin{subequations}
\label{eq:acousticendenReq}
\begin{align}
\partial_\rmx u_\rmz^\eff &= f \partial_\rmx u_\rmz + (1-f) \partial_\rmx u_\rmz^\prime, \\
 \sigma_\xx^\eff &= f \sigma_\xx + (1-f) \sigma_\xx^\prime, \\
\partial_\rmy u_\rmz^\eff &= f \partial_\rmy u_\rmz + (1-f) \partial_\rmy u_\rmz^\prime, \\
 \sigma_\yy^\eff &= f \sigma_\yy + (1-f) \sigma_\yy^\prime, \\
\partial_\rmz u_\rmz^\eff &= f \partial_\rmz u_\rmz + (1-f) \partial_\rmz u_\rmz^\prime, \\
 \sigma_\xy^\eff &= f \sigma_\xy + (1-f) \sigma_\xy^\prime, 
\end{align}
\end{subequations}
where for convenience, we now differentiate \eqref{eq:contdisptrans} and compile these with the derivatives of  the displacement fields in \eqref{eq:acousticendenReq} along with the stress fields to obtain
\begin{equation}
\label{eq:acCONDS}
\begin{array}{lcl}
\sigma_\xx^\eff = f \sigma_\xx + (1-f) \sigma_\xx^\prime, &  \quad&   s_\xx^\eff = s_\xx = s_\xx^\prime, \\
\sigma_\yy^\eff = f \sigma_\yy + (1-f) \sigma_\yy^\prime, &    \quad&    s_\yy^\eff = s_\yy = s_\yy^\prime, \\
\sigma_\zz^\eff = \sigma_\zz= \sigma_\zz^\prime, 		&   \quad&     s_\zz^\eff = f s_\zz + (1-f) s_\zz^\prime, \\
\sigma_\yz^\eff = \sigma_\yz= \sigma_\yz^\prime, 		&    \quad&    s_\yz^\eff = f s_\yz + (1-f) s_\yz^\prime, \\
\sigma_\xz^\eff = \sigma_\xz= \sigma_\xz^\prime, 		&  \quad&   s_\xz^\eff = f s_\xz + (1-f) s_\xz^\prime, \\
\sigma_\xy^\eff = f \sigma_\xy + (1-f) \sigma_\xy^\prime,       &  \quad& s_\xy^\eff = s_\xy = s_\xy^\prime. 
 \end{array}
\end{equation}
Using     \eqref{eq:acCONDS} along with the constitutive relations
\begin{equation}
\label{eq:mechCONST}
\sigma_{ij}^\eff = C_{ijkl}^\eff s_{kl}^\eff, \;  \sigma_{ij}  = C_{ijkl}  s_{kl}, \; \sigma_{ij}^\prime = C_{ijkl}^\prime s_{kl}^\prime,
\end{equation}
where $C_{ijkl}$ denotes the linear stiffness tensor, we recover the effective stiffness coefficients. Here the layers comprise optically isotropic  media, from which we obtain all six  unique non-vanishing parameters of the mechanical stiffness tensor   $C_{ijkl}^\eff$ for an effective tetragonal ($4/mmm$) material \cite{nye1985physical} as
\begin{subequations}
\label{eq:allCeff}
\begin{multline}
C_{\xx\xx}^\eff    = f  C_{\xx\xx}   + (1-f)   C_{\xx\xx}^\prime \\
- \frac{f(1-f) (C_{\xx\yy} - C_{\xx\yy}^\prime)^2}{ f C_{\xx\xx}^\prime + (1-f) C_{\xx\xx}}   
\end{multline}
%\vspace{-7mm}	%unnecessary spacing between multlines
 \begin{multline}
C_{\xx\yy}^\eff    = f  C_{\xx\yy}   + (1-f)   C_{\xx\yy}^\prime  \\
- \frac{f(1-f) (C_{\xx\yy} - C_{\xx\yy}^\prime)^2}{ f C_{\xx\xx}^\prime + (1-f) C_{\xx\xx}}   
\end{multline}
%\vspace{-4mm}
\begin{align}
 C_{\xx\zz}^\eff  &=  \frac{f C_{\xx\yy} C_{\xx\xx}^\prime  +  (1-f) C_{\xx\xx} C_{\xx\yy}^\prime  }{ f  C_{\xx\xx}^\prime + (1-f) C_{\xx\xx}   },  \\
 \frac{1}{C_{\zz\zz}^\eff}  &=  \frac{f}{C_{\xx\xx}} +  \frac{(1-f)}{C_{\xx\xx}^\prime}, \\
\frac{1}{C_{\yz\yz}^\eff}  &=  \frac{f}{C_{\yz\yz}} +  \frac{(1-f)}{C_{\yz\yz}^\prime}, \\
C_{\xy\xy}^\eff &= f C_{\yz\yz}  +(1-f)  C_{\yz\yz}^\prime.
\end{align}
\end{subequations}
The expressions in \eqref{eq:allCeff} above  are equivalent to those presented in \citet{rouhani1986effective} and  \citet{grimsditch1985effective}, after considering the symmetry properties of the constituent layers.

\subsection{Effective symmetric photoelastic tensor} \label{sec:pijkleff} \noindent
In this section we evaluate the symmetric     photoelastic tensor $p_{ijkl}^\eff  $    defined by
\begin{subequations}
\begin{align}
\Delta (\varepsilon_\eff^{ -1})_{ij} &= p_{ijkl}^\eff s_{kl}^\eff, \qquad \qquad \mbox{ or equivalently, } \\
\label{eq:photoeleffDEF}
 \Delta (\varepsilon_\eff)_{ij} &= -\varepsilon_{ii}^\eff \, \varepsilon_{jj}^\eff \, p_{ijkl}^\eff \, s_{kl}^\eff,
\end{align}
\end{subequations}
provided the medium does not possess triclinic or monoclinic symmetry\cite{sapriel1979acousto}.   Expressions for the effective photoelastic tensor elements $p_{ijkl}^\eff$ are obtained by differentiating  the  effective permittivity expressions $\varepsilon_{ij}^\eff$ in \eqref{eq:effeps} with respect to  individual strain fields   $s_{ij}^\eff$ defined on the  length scale of the unit cell whilst holding other effective strain fields constant. For example,   \eqref{eq:photoeleffDEF} for an effective tetragonal $(4/mmm)$ material is given by
\begin{equation}
\label{eq:photoelepszzrow}
\Delta \varepsilon_{\zz}^\eff = - \left( \varepsilon_\zz^\eff \right)^2 \, \left[ p_{\zz\xx}^\eff    s_{\xx}^\eff + p_{\zz\xx}^\eff    s_\yy^\eff  + p_{\zz\zz}^\eff    s_{\zz}^\eff \right],
\end{equation}
and subsequently
\begin{subequations}
\label{eq:derivepszzdefn}
\begin{align}
\label{eq:derivepszzesxxedefn}
\frac{\partial \varepsilon_\zz^\eff}{\partial s_\xx^\eff}\bigg|_{s_\yy^\eff,s_\zz^\eff} = - \left( \varepsilon_\zz^\eff \right)^2  p_{\zz\xx}^\eff  , \\
\label{eq:derivepszzesyyedefn}
\frac{\partial \varepsilon_\zz^\eff}{\partial s_\yy^\eff}\bigg|_{s_\xx^\eff,s_\zz^\eff} = - \left( \varepsilon_\zz^\eff \right)^2  p_{\zz\xx}^\eff  , \\
\label{eq:derivepszzeszzedefn}
\frac{\partial \varepsilon_\zz^\eff}{\partial s_\zz^\eff}\bigg|_{s_\xx^\eff,s_\yy^\eff} = - \left( \varepsilon_\zz^\eff \right)^2  p_{\zz\zz}^\eff.
\end{align}
\end{subequations}
Therefore, an analytical expression for $p_{\zz\zz}^\eff$ is recovered by differentiating  $\varepsilon_\zz^\eff$   \eqref{eq:effepszz} with respect to $s_{\zz}^\eff$,  with both $s_\xx^\eff$ and $s_\yy^\eff$ held constant. The resulting expressions are then  reduced   using the tensor definitions for the constituent materials
\begin{subequations}
\label{eq:allphotoels}
\begin{align}
\label{eq:photoelmat1DEF}
\Delta \varepsilon_{ij} &= -\varepsilon_{ii} \, \varepsilon_{jj} \, p_{ijkl} \, s_{kl}, \\
\label{eq:photoelmat2DEF}
\Delta \varepsilon_{ij}^\prime &= -\varepsilon_{ii}^\prime \, \varepsilon_{jj}^\prime \, p_{ijkl}^\prime \, s_{kl}^\prime,
\end{align}
\end{subequations}
the mechanical constitutive relations \eqref{eq:mechCONST}, and the relationships between stress and strain fields in \eqref{eq:acCONDS}. We remark that the permittivities $\varepsilon_{ij}$ and $\varepsilon_{ij}^\prime$ are functions of their constituent strain fields  alone (see Eq. \eqref{eq:allphotoels}).   The derivation for all seven unique non-vanishing photoelastic constants necessary to describe a layered structure is extensive and we feel that there is little merit in providing  a    complete outline for all  terms. Accordingly, we   consider  the derivations for $p_{\zz\xx}^\eff$ and $p_{\zz\zz}^\eff$  alone and present the final expressions for all remaining coefficients.  

 As identified in \eqref{eq:derivepszzesxxedefn}  above, we now implicitly differentiate the effective permittivity expression \eqref{eq:effepszz} with respect to $s_\xx^\eff$, holding the strain fields $s_\yy^\eff$ and $s_\zz^\eff$ constant, which admits
\begin{multline}
\label{eq:derivepszzesxxe}
 \frac{1}{(\varepsilon_\zz^\eff)^2} \frac{\partial \varepsilon_\zz^\eff}{\partial s_\xx^\eff}\bigg|_{s_\yy^\eff,s_\zz^\eff}  = 
 \frac{f}{(\varepsilon_\zz)^2} \frac{\partial \varepsilon_\zz}{\partial s_\xx^\eff}\bigg|_{s_\yy^\eff,s_\zz^\eff}  \\
 \frac{(1-f)}{(\varepsilon_\zz^\prime)^2} \frac{\partial \varepsilon_\zz^\prime}{\partial s_\xx^\eff} \bigg|_{s_\yy^\eff,s_\zz^\eff} 
- \left( \frac{1}{\varepsilon_\zz} - \frac{1}{\varepsilon_\zz^\prime} \right) \frac{\partial f}{\partial s_\xx^\eff}\bigg|_{s_\yy^\eff,s_\zz^\eff}.
\end{multline}
The first derivative in \eqref{eq:derivepszzesxxe} follows immediately from the definition in \eqref{eq:derivepszzesxxedefn} above. The next derivative is evaluated using the definition of the photoelastic tensor \eqref{eq:photoelmat1DEF} in the first optically isotropic layer.  An application of chain rule then admits
\begin{multline}
\label{eq:firstderivepszz}
 \frac{\partial \varepsilon_\zz}{\partial s_\xx^\eff} \bigg|_{s_\yy^\eff,s_\zz^\eff}= 
 \frac{\partial \varepsilon_\zz}{\partial s_\xx }\bigg|_{s_\yy^\eff,s_\zz^\eff} \frac{\partial s_\xx }{\partial s_\xx^\eff} \bigg|_{s_\yy^\eff,s_\zz^\eff}  \\ +
 \frac{\partial \varepsilon_\zz}{\partial s_\yy}\bigg|_{s_\yy^\eff,s_\zz^\eff} \frac{\partial s_\yy}{\partial s_\xx^\eff} \bigg|_{s_\yy^\eff,s_\zz^\eff}+ 
 \frac{\partial \varepsilon_\zz}{\partial s_\zz}\bigg|_{s_\yy^\eff,s_\zz^\eff} \frac{\partial s_\zz}{\partial s_\xx^\eff}\bigg|_{s_\yy^\eff,s_\zz^\eff}.
\end{multline}
Using the acoustic boundary conditions $s_\xx= s_\xx^\eff $ and $s_\yy = s_\yy^\eff$ from \eqref{eq:acCONDS} we have that
\begin{equation}
\label{eq:first2sxxa}
\frac{\partial s_\xx }{\partial s_\xx^\eff} \bigg|_{s_\yy^\eff,s_\zz^\eff} = 1, \quad \mbox{ and} \quad  \frac{\partial s_\yy }{\partial s_\xx^\eff} \bigg|_{s_\yy^\eff,s_\zz^\eff} =0,
\end{equation}
respectively. The boundary condition $\sigma_\zz = \sigma_\zz^\eff$ and the constitutive relations  for the constituent layers \eqref{eq:mechCONST}  give
\begin{multline}
\label{eq:equateHookes1}
C_{\xx\zz}^\eff  s_\xx^\eff + C_{\xx\zz}^\eff s_\yy^\eff  +  C_{\zz\zz}^\eff s_\zz^\eff =  \\
C_{\xx\yy}   s_\xx  + C_{\xx\yy}  s_\yy  +  C_{\xx\xx}  s_\zz ,
\end{multline}
which after implicit differentiation, where we also hold the strain fields $s_\yy^\eff$ and $s_\zz^\eff$ constant, takes the form
\begin{subequations}
\begin{multline}
\label{eq:equatederivHookes1}
C_{\xx\zz}^\eff  \frac{\partial s_\xx^\eff}{\partial s_\xx^\eff}\bigg|_{s_\yy^\eff,s_\zz^\eff} + C_{\xx\zz}^\eff \frac{\partial s_\yy^\eff}{\partial s_\xx^\eff}\bigg|_{s_\yy^\eff,s_\zz^\eff}  \\
+  C_{\zz\zz}^\eff \frac{\partial s_\zz^\eff}{\partial s_\xx^\eff}\bigg|_{s_\yy^\eff,s_\zz^\eff}     =
C_{\xx\yy}   \frac{\partial s_\xx}{\partial s_\xx^\eff} \bigg|_{s_\yy^\eff,s_\zz^\eff} \\
+ C_{\xx\yy}  \frac{\partial s_\yy}{\partial s_\xx^\eff}\bigg|_{s_\yy^\eff,s_\zz^\eff} +  C_{\xx\xx}  \frac{\partial s_\zz}{\partial s_\xx^\eff}\bigg|_{s_\yy^\eff,s_\zz^\eff}.
\end{multline}
The   boundary conditions \eqref{eq:acCONDS} and   constant field requirements evident  from \eqref{eq:derivepszzesxxedefn} reduce    \eqref{eq:equatederivHookes1} to the form
\begin{equation}
C_{\xx\zz}^\eff     = 
C_{\xx\yy}   +  C_{\xx\xx}   \frac{\partial s_\zz}{\partial s_\xx^\eff}\bigg|_{s_\yy^\eff,s_\zz^\eff},
\end{equation}
and ultimately admits   
\begin{equation}
\label{eq:secondsxx}
\frac{\partial s_\zz}{\partial s_\xx^\eff} \bigg|_{s_\yy^\eff,s_\zz^\eff} = \frac{C_{\xx\zz}^\eff - C_{\xx\zz} }{C_{\xx\xx}}.
\end{equation}
\end{subequations}
Substituting \eqref{eq:first2sxxa} and \eqref{eq:secondsxx} into the derivative \eqref{eq:firstderivepszz} gives
\begin{subequations}
\begin{equation}
\label{eq:epszzosxxe}
 \frac{\partial \varepsilon_\zz}{\partial s_\xx^\eff} \bigg|_{s_\yy^\eff,s_\zz^\eff}= -(\varepsilon_\zz)^2 p_{\xx\zz} -(\varepsilon_\zz)^2 p_{\xx\xx} \left[ \frac{C_{\xx\zz}^\eff - C_{\xx\zz} }{C_{\xx\xx}} \right],
\end{equation}
after using \eqref{eq:photoelmat1DEF}, and analogously we have that 
\begin{equation}
\label{eq:epszzposxxe}
 \frac{\partial \varepsilon_\zz^\prime}{\partial s_\xx^\eff} \bigg|_{s_\yy^\eff,s_\zz^\eff}= -(\varepsilon_\zz^\prime)^2 p_{\xx\zz}^\prime -(\varepsilon_\zz^\prime)^2 p_{\xx\xx}^\prime \left[ \frac{C_{\xx\zz}^\eff - C_{\xx\zz}^\prime }{C_{\xx\xx}^\prime} \right],
\end{equation}
\end{subequations}
for the second optically isotropic layer, following the boundary condition $\sigma_\zz^\prime = \sigma_\zz^\eff$ \eqref{eq:acCONDS}.
 
 The derivative of the filling fraction in \eqref{eq:derivepszzesxxe}, through an application of chain rule, gives rise to  
 \begin{subequations}
 \begin{multline}
\frac{\partial f}{\partial s_\xx^\eff}\bigg|_{s_\yy^\eff,s_\zz^\eff} 
 = a \frac{\partial f}{\partial a}\bigg|_{s_\yy^\eff,s_\zz^\eff,a^\prime} \frac{\partial s_\zz}{\partial s_\xx^\eff}\bigg|_{s_\yy^\eff,s_\zz^\eff}    \\
 +a^\prime
\frac{\partial f}{\partial a^\prime}\bigg|_{s_\yy^\eff,s_\zz^\eff,a} \frac{\partial s_\zz^\prime}{\partial s_\xx^\eff}\bigg|_{s_\yy^\eff,s_\zz^\eff},
 \end{multline}
following from the definitions  $f = a/(a+a^\prime)$, $\Delta s_\zz = \Delta a/ a$ and   $\Delta s_\zz^\prime = \Delta a^\prime / a^\prime$. Using \eqref{eq:secondsxx} and the analogous expression for the second layer, we obtain
\begin{equation}
\label{eq:fosxxe}
\frac{\partial f}{\partial s_\xx^\eff}\bigg|_{s_\yy^\eff,s_\zz^\eff}  = f(1-f) \left[ \frac{C_{\xx\zz}^\eff - C_{\xx\yy} }{C_{\xx\xx} } - \frac{C_{\xx\zz}^\eff - C_{\xx\yy}^\prime }{C_{\xx\xx}^\prime } \right].
\end{equation}
\end{subequations}
Substituting \eqref{eq:derivepszzesxxedefn}, \eqref{eq:epszzosxxe}, \eqref{eq:epszzposxxe}, and \eqref{eq:fosxxe} into the expression \eqref{eq:derivepszzesxxe} gives
\begin{multline}
\label{eq:pzzxxefforig}
p_{\zz\xx}^\eff = f p_{\xx\yy} + (1-f) p_{\xx\yy}^\prime \\
- \frac{f(1-f) (p_{\xx\xx} - p_{\xx\xx}^\prime) (C_{\xx\yy} - C_{\xx\yy}^\prime )}{ f C_{\xx\xx}^\prime + (1-f) C_{\xx\xx}} \\
- f(1-f) \left( \frac{1}{\varepsilon_\zz}  - \frac{1}{\varepsilon_\zz^\prime} \right) \left( \frac{C_{\xx\yy} - C_{\xx\yy}^\prime }{ f C_{\xx\xx}^\prime + (1-f) C_{\xx\xx}} \right).
\end{multline}
Having determined the analytical expression for $p_{\zz\xx}^\eff$, we now turn to the derivation of the $p_{\zz\zz}^\eff$ coefficient. 
Similarly, implicit differentiation of the effective permittivity expression \eqref{eq:effepszz} with respect to $s_\zz^\eff$, with $s_\xx^\eff$ and $s_\yy^\eff$ held constant, admits
 \begin{multline}
\label{eq:derivepszzeszze}
-\frac{1}{(\varepsilon_\zz^\eff)^2} \frac{\partial \varepsilon_\zz^\eff}{\partial s_\zz^\eff}\bigg|_{s_\xx^\eff,s_\yy^\eff}  = 
-\frac{f}{(\varepsilon_\zz)^2} \frac{\partial \varepsilon_\zz}{\partial s_\zz^\eff}\bigg|_{s_\xx^\eff,s_\yy^\eff} \\
-\frac{(1-f)}{(\varepsilon_\zz^\prime)^2} \frac{\partial \varepsilon_\zz^\prime}{\partial s_\zz^\eff} \bigg|_{s_\xx^\eff,s_\yy^\eff} 
+ \left( \frac{1}{\varepsilon_\zz} - \frac{1}{\varepsilon_\zz^\prime} \right) \frac{\partial f}{\partial s_\zz^\eff}\bigg|_{s_\xx^\eff,s_\yy^\eff},
\end{multline}
where the first derivative is given by \eqref{eq:derivepszzeszzedefn}. In an analogous procedure to before, we have that
 \begin{multline}
\label{eq:firstderivepszzb}
 \frac{\partial \varepsilon_\zz}{\partial s_\zz^\eff} \bigg|_{s_\xx^\eff,s_\yy^\eff}= 
 \frac{\partial \varepsilon_\zz}{\partial s_\xx }\bigg|_{s_\xx^\eff,s_\yy^\eff} \frac{\partial s_\xx }{\partial s_\zz^\eff} \bigg|_{s_\xx^\eff,s_\yy^\eff}   \\
 + \frac{\partial \varepsilon_\zz}{\partial s_\yy}\bigg|_{s_\xx^\eff,s_\yy^\eff} \frac{\partial s_\yy}{\partial s_\zz^\eff} \bigg|_{s_\xx^\eff,s_\yy^\eff}+ 
 \frac{\partial \varepsilon_\zz}{\partial s_\zz}\bigg|_{s_\xx^\eff,s_\yy^\eff} \frac{\partial s_\zz}{\partial s_\zz^\eff}\bigg|_{s_\xx^\eff,s_\yy^\eff},
\end{multline}
where the  boundary conditions \eqref{eq:acCONDS} and constant field requirements give rise to
\begin{equation}
\label{eq:first2sxx}
\frac{\partial s_\xx }{\partial s_\zz^\eff} \bigg|_{s_\xx^\eff,s_\yy^\eff} =   \frac{\partial s_\yy }{\partial s_\zz^\eff} \bigg|_{s_\yy^\eff,s_\zz^\eff} =0.
\end{equation}
The remaining boundary condition $\sigma_\zz = \sigma_\zz^\eff$ and constitutive relations \eqref{eq:mechCONST} give the expression \eqref{eq:equateHookes1} once more. Implicit differentiation  with respect to $s_\zz^\eff$ and the  new constant field requirements  admits
\begin{multline}
\label{eq:equatederivHookes2}
C_{\xx\zz}^\eff  \frac{\partial s_\xx^\eff}{\partial s_\zz^\eff}\bigg|_{s_\xx^\eff,s_\yy^\eff} + C_{\xx\zz}^\eff \frac{\partial s_\yy^\eff}{\partial s_\zz^\eff}\bigg|_{s_\xx^\eff,s_\yy^\eff} \\
 +  C_{\zz\zz}^\eff \frac{\partial s_\zz^\eff}{\partial s_\zz^\eff}\bigg|_{s_\xx^\eff,s_\yy^\eff} = 
C_{\xx\yy}   \frac{\partial s_\xx}{\partial s_\zz^\eff}\bigg|_{s_\xx^\eff,s_\yy^\eff} \\
 + C_{\xx\yy}  \frac{\partial s_\yy}{\partial s_\zz^\eff}\bigg|_{s_\xx^\eff,s_\yy^\eff} +  C_{\xx\xx}  \frac{\partial s_\zz}{\partial s_\zz^\eff}\bigg|_{s_\xx^\eff,s_\yy^\eff},
\end{multline}
and ultimately  we find that
\begin{equation}
\label{eq:secondszzsxx}
\frac{\partial s_\zz}{\partial s_\zz^\eff}\bigg|_{s_\xx^\eff,s_\yy^\eff} = \frac{C_{\zz\zz}^\eff}{C_{\xx\xx}}.
\end{equation}
Subsequently, substituting \eqref{eq:first2sxx} and \eqref{eq:secondszzsxx} into \eqref{eq:firstderivepszzb} we obtain
\begin{subequations}
\begin{equation}
\label{eq:epszzecond1}
 \frac{\partial \varepsilon_\zz}{\partial s_\zz^\eff} \bigg|_{s_\xx^\eff,s_\yy^\eff}= 
 -(\varepsilon_\zz)^2 p_{\xx\xx} \left( \frac{C_{\zz\zz}^\eff}{C_{\xx\xx}} \right),
\end{equation}
after using \eqref{eq:photoelmat1DEF}, and analogously we have
\begin{equation}
\label{eq:epszzecond2}
 \frac{\partial \varepsilon_\zz^\prime}{\partial s_\zz^\eff} \bigg|_{s_\xx^\eff,s_\yy^\eff}= 
 -(\varepsilon_\zz^\prime)^2 p_{\xx\xx}^\prime \left( \frac{C_{\zz\zz}^\eff}{C_{\xx\xx}^\prime} \right).
\end{equation}
\end{subequations}
The derivative of the filling fraction in \eqref{eq:derivepszzeszze}   takes the form
\begin{multline}
\label{eq:epszzecond3}
 \frac{\partial f}{\partial s_\zz^\eff}\bigg|_{s_\xx^\eff,s_\yy^\eff} = f(1-f) \left[ \frac{\partial s_\zz}{\partial s_\zz^\eff}\bigg|_{s_\xx^\eff,s_\yy^\eff} - \frac{\partial s_\zz^\prime}{\partial s_\zz^\eff} \bigg|_{s_\xx^\eff,s_\yy^\eff}\right] \\
 = f(1-f) \left[ \frac{C_{\zz\zz}^\eff}{C_{\xx\xx}} - \frac{C_{\zz\zz}^\eff}{C_{\xx\xx}^\prime} \right],
\end{multline}
 following \eqref{eq:secondszzsxx} and the corresponding expression for the second layer. Substituting \eqref{eq:epszzecond1}, \eqref{eq:epszzecond2}, and \eqref{eq:epszzecond3} into \eqref{eq:derivepszzeszze} we obtain
 \begin{multline}
 \label{eq:pzzzzefffirstobs}
 \frac{p_{\zz\zz}^\eff}{C_{\zz\zz}^\eff} = f \left(\frac{p_{\xx\xx} }{C_{\xx\xx} } \right) +(1- f) \left(\frac{p_{\xx\xx}^\prime }{C_{\xx\xx}^\prime } \right) \\
 + f(1-f) \left( \frac{1}{\varepsilon_\zz} - \frac{1}{\varepsilon_\zz^\prime} \right) \left( \frac{1}{C_{\xx\xx} } - \frac{1}{C_{\xx\xx}^\prime} \right)
 \end{multline}
The expressions for $p_{\zz\xx}^\eff$ in \eqref{eq:pzzxxefforig} and $p_{\zz\zz}^\eff$ in \eqref{eq:pzzzzefffirstobs} are  presented below in \eqref{eq:allPIJKLEff} along with all other remaining coefficients  
\begin{subequations}
\label{eq:allPIJKLEff}
\begin{widetext}
\begin{align}
(\varepsilon_\xx^\eff)^2 p_{\xx\xx}^\eff &= f(\varepsilon_\xx )^2 p_{\xx\xx} + (1-f) (\varepsilon_\xx^\prime)^2 p_{\xx\xx}^\prime \\ \nonumber &\quad\shoveright{
- \frac{ f(1-f) (C_{\xx\yy} - C_{\xx\yy}^\prime) \left[ (\varepsilon_\xx )^2 p_{\xx\yy} - (\varepsilon_\xx^\prime )^2 p_{\xx\yy}^\prime \right]}{f C_{\xx\xx}^\prime + (1-f) C_{\xx\xx} } 
+ \frac{f(1-f) (\varepsilon_\xx - \varepsilon_\xx^\prime) (C_{\xx\yy} - C_{\xx\yy}^\prime)}{ f C_{\xx\xx}^\prime + (1-f) C_{\xx\xx}},} \\
 \frac{p_{\zz\zz}^\eff}{C_{\zz\zz}^\eff} &= f \left(\frac{p_{\xx\xx} }{C_{\xx\xx} } \right) +(1- f) \left(\frac{p_{\xx\xx}^\prime }{C_{\xx\xx}^\prime } \right) + f(1-f) \left( \frac{1}{\varepsilon_\zz} - \frac{1}{\varepsilon_\zz^\prime} \right) \left( \frac{1}{C_{\xx\xx} } - \frac{1}{C_{\xx\xx}^\prime} \right),\\
\frac{(\varepsilon_\xx^\eff)^2 p_{\xx\zz}^\eff}{C_{\zz\zz}^\eff} &= f \frac{(\varepsilon_\xx)^2 p_{\xx\yy}}{C_{\xx\xx}} + (1-f)\frac{(\varepsilon_\xx^\prime)^2 p_{\xx\yy}^\prime}{C_{\xx\xx}^\prime} - f(1-f)(\varepsilon_\xx - \varepsilon_\xx^\prime)\left( \frac{1}{C_{\xx\xx} } - \frac{1}{C_{\xx\xx}^\prime}\right), \\
(\varepsilon_\xx^\eff)^2 p_{\xx\yy}^\eff &= f(\varepsilon_\xx )^2 p_{\xx\yy} + (1-f) (\varepsilon_\xx^\prime)^2 p_{\xx\yy}^\prime  \\ \nonumber&\quad\shoveright{
- \frac{ f(1-f) (C_{\xx\yy} - C_{\xx\yy}^\prime) \left[ (\varepsilon_\xx )^2 p_{\xx\yy} - (\varepsilon_\xx^\prime )^2 p_{\xx\yy}^\prime \right]}{f C_{\xx\xx}^\prime + (1-f) C_{\xx\xx} } 
+ \frac{f(1-f) (\varepsilon_\xx - \varepsilon_\xx^\prime) (C_{\xx\yy} - C_{\xx\yy}^\prime)}{ f C_{\xx\xx}^\prime + (1-f) C_{\xx\xx}},}\\
 p_{\zz\xx}^\eff &= f p_{\xx\yy} + (1-f) p_{\xx\yy}^\prime \\ \nonumber &\quad\shoveright{
 - \frac{f(1-f) (p_{\xx\xx} - p_{\xx\xx}^\prime) (C_{\xx\yy} - C_{\xx\yy}^\prime )}{ f C_{\xx\xx}^\prime + (1-f) C_{\xx\xx}} - f(1-f) \left( \frac{1}{\varepsilon_\zz}  - \frac{1}{\varepsilon_\zz^\prime} \right) \left( \frac{C_{\xx\yy} - C_{\xx\yy}^\prime }{ f C_{\xx\xx}^\prime + (1-f) C_{\xx\xx}} \right),} \\
\frac{\varepsilon_\yy^\eff p_{\yz\yz}^\eff}{C_{\yz\yz}^\eff} &=
f \frac{\varepsilon_\yy p_{\yz\yz}}{C_{\yz\yz}} + 
(1-f) \frac{\varepsilon_\yy^\prime p_{\yz\yz}^\prime}{C_{\yz\yz}^\prime}, \\
 \varepsilon_\xx^\eff \varepsilon_\yy^\eff p_{\xy\xy}^\eff &= f   \varepsilon_\xx \varepsilon_\yy p_{\xy\xy} + (1- f)   \varepsilon_\xx^\prime \varepsilon_\yy^\prime p_{\xy\xy}^\prime .
\end{align}
\end{widetext}
\end{subequations}
We remark that in the expressions above, the photoelastic ocefficients possess the form $p_{ijkl}^\eff = \alpha_{qrst} \, p_{qrst} + \alpha^\prime_{qrst} \, p_{qrst}^\prime + p_{ijkl}^\art$, where $\alpha_{qrst}$ and $\alpha^\prime_{qrst}$ are functions of material parameters, but may be regarded as   weightings for the photoelastic coefficients of the constituent layers. Following the convention established in earlier work \cite{smith2015electrostriction}, the final contribution $p_{ijkl}^\art$ is termed  artificial photoelasticity, as this represents a non-trivial contribution to the photoelastic properties of the composite when $p_{qrst} = p_{qrst}^\prime = 0$. These artificial contributions are  directly proportional to the contrast in relevant components of the permittivity and stiffness tensors,  and   have been shown to play a significant role in the photoelastic properties of other subwavelength structured designs \cite{smith2016control,smith2016stimulated}. Note that  for  $p_{\yz\yz}^\eff$ and $p_{\xy\xy}^\eff$ above, there is no artificial photoelastic component,    as shear waves do not change the volume of the unit cell when the constituent and effective material are oriented with the Cartesian coordinate frame, i.e.,
\begin{equation}
\frac{\partial f}{\partial s_{\yz}^\eff}  =  \quad \frac{\partial f}{\partial s_{\xz}^\eff} =  \quad \frac{\partial f}{\partial s_{\xy}^\eff} = 0,
\end{equation}
however, we emphasize that this result only holds for high-symmetry composites.

The derivation outlined in this section gives results for the symmetric  photoelastic {\it  strain} tensor, and that expressions for the symmetric  photoelastic {\it  stress}  tensor may be  found through a straightforward application of Hooke's law \cite{nye1985physical}. However,      the photoelastic strain coefficients $p_{ijkl}^\eff$  may   be {\it expressed} in terms of    acoustic fields that are everywhere continuous in the layered medium,  analogously to   \citet{rouhani1986effective}. For example,     substituting the constitutive relation \eqref{eq:mechCONST}  into the photoelastic tensor definition \eqref{eq:photoelepszzrow} we obtain
 \begin{multline}
 \Delta \varepsilon_{\zz}^\eff = - \left( \varepsilon_\zz^\eff \right)^2 \, \left[  \left\{ p_{\zz\xx}^\eff   - \frac{C_{\xx\zz}^\eff}{C_{\zz\zz}^\eff} p_{\zz\zz}^\eff\right\}  s_{\xx}^\eff   \right.\\ \left.
 + \left\{ p_{\zz\xx}^\eff   - \frac{C_{\xx\zz}^\eff}{C_{\zz\zz}^\eff} p_{\zz\zz}^\eff\right\} s_\yy^\eff  + \frac{ p_{\zz\zz}^\eff}{C_{\zz\zz}^\eff}    \sigma_{\zz}^\eff \right],
 \end{multline}
 where photoelastic strain coefficients are now obtained through differentiation (as before), but with effective stress and effective strain fields held constant. However,   such a procedure gives identical final expressions for $p_{ijkl}^\eff$ to those presented in \eqref{eq:allPIJKLEff}. As a final remark, the   $p_{ijkl}^\eff$ presented   in \eqref{eq:allPIJKLEff}, with $p_{ijkl}^\art = 0$, are identical to those tabulated in  \citet{rouhani1986effective} after considering symmetry reductions  of   tensor coefficients\cite{nye1985physical}.

\subsection{Effective antisymmetric photoelastic tensor}\label{sec:pijklasymmeff} \noindent
In this section,  we evaluate the antisymmetric component of the photoelastic tensor $r_{ijkl}^\eff$ defined by
\begin{equation}
\Delta(\varepsilon^{-1}_\eff)_{ij} = r_{ijkl}^\eff r_{kl}^\eff,
\end{equation}
where $r_{kl} = \tfrac{1}{2}(\partial_l u_k - \partial_k u_l)$ denotes the infinitesimal rotation tensor. The derivation for the roto-optic tensor of   a uniform material   is given in \cite{nelson1970new,nelson1979electric} and extends   to the case of a subwavelength structured material as
\begin{subequations}
\begin{multline}
r_{ijkl}^\eff  = \frac{1}{2} \left[ (\varepsilon^{-1}_\eff)_{il} \delta_{kj}  + (\varepsilon^{-1}_\eff)_{lj} \delta_{ik} \right. \\
\left. - (\varepsilon^{-1}_\eff)_{ik} \delta_{lj} - (\varepsilon^{-1}_\eff)_{kj} \delta_{il} \right].
\end{multline}
This simplifies to the form \cite{sapriel1979acousto}
\begin{equation}
 r_{ij kl }^\eff = \frac{1}{2}  \left( \frac{1}{\varepsilon_{jj}^\eff} -  \frac{1}{\varepsilon_{ii}^\eff} \right) (\delta_{ik} \delta_{jl} - \delta_{il} \delta_{jk} ) ,
 \end{equation}
\end{subequations}
 provided the layered material does not possess triclinic or monoclinic symmetry. For our tetragonal ($4/mmm$) layered structure, there are only eight non-vanishing $r_{ijkl}$ terms, which all take  the same value modulo a sign change that arises from the antisymmetric nature of the tensor $r_{ijkl} =-r_{ijlk}$.
From the expressions for the effective permittivity presented in Section \ref{sec:effperm}, we have that  
\begin{equation}
r_{\xz \xz }^\eff   = \frac{1}{2} \left( \frac{f \varepsilon_\zz^\prime + (1-f) \varepsilon_\zz}{\varepsilon_\zz \varepsilon_\zz^\prime} - \frac{1}{f \varepsilon_\xx + (1-f) \varepsilon_\xx^\prime} \right),
\end{equation}
 for layers of optically isotropic media.
 
 \section{Numerical Examples} \label{sec:numerexamp} \noindent
  
  In this section, we present the effective permittivity, stiffness, and photoelastic tensors for a selection of   material combinations, where constituent  parameter values are taken from Table 1 of \citet{smith2016control}. Here values are presented at a vacuum wavelength of $\lambda = 1550 \, \nm$ and for a total layer cell width of $a + a^\prime = 50 \, \nm$.  
   
\begin{center}
 \begin{figure}[t]
\subfigure[ \label{fig:sisio2p1}]{
\includegraphics[scale=0.28]{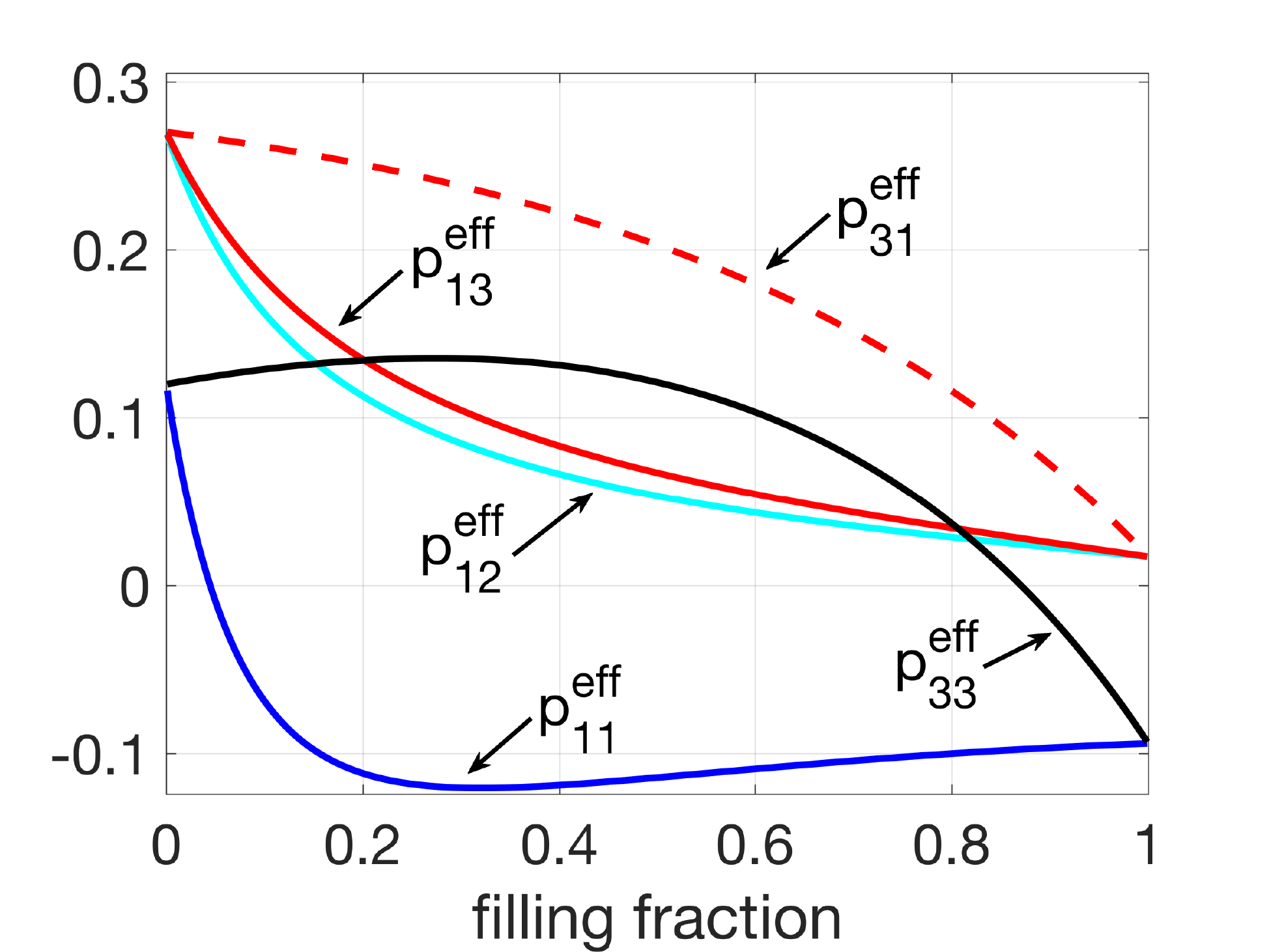}}
\subfigure[  \label{fig:sisio2p2} ]{
\includegraphics[scale=0.28]{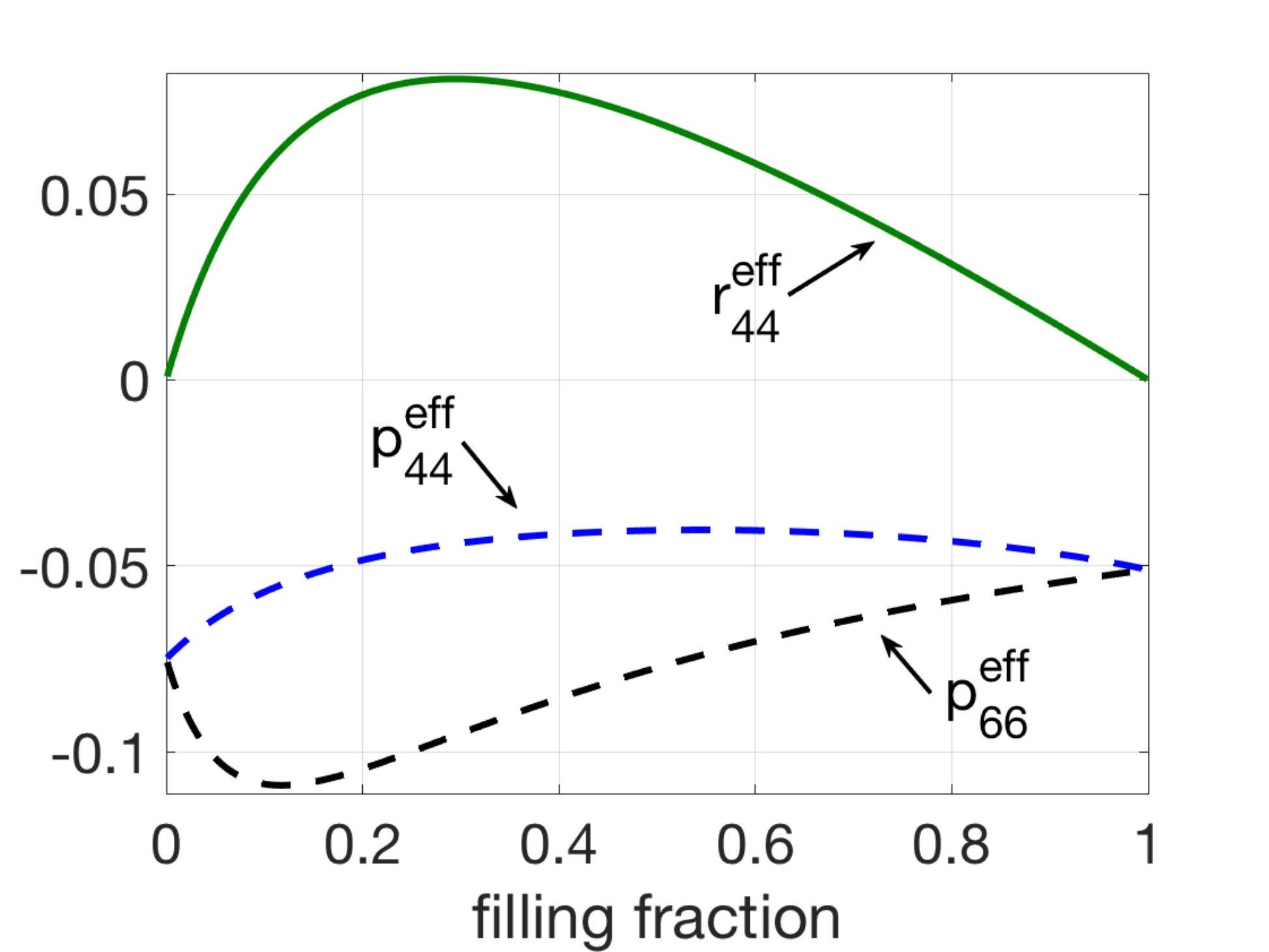}}
\subfigure[  \label{fig:sisio2art} ]{
\includegraphics[scale=0.28]{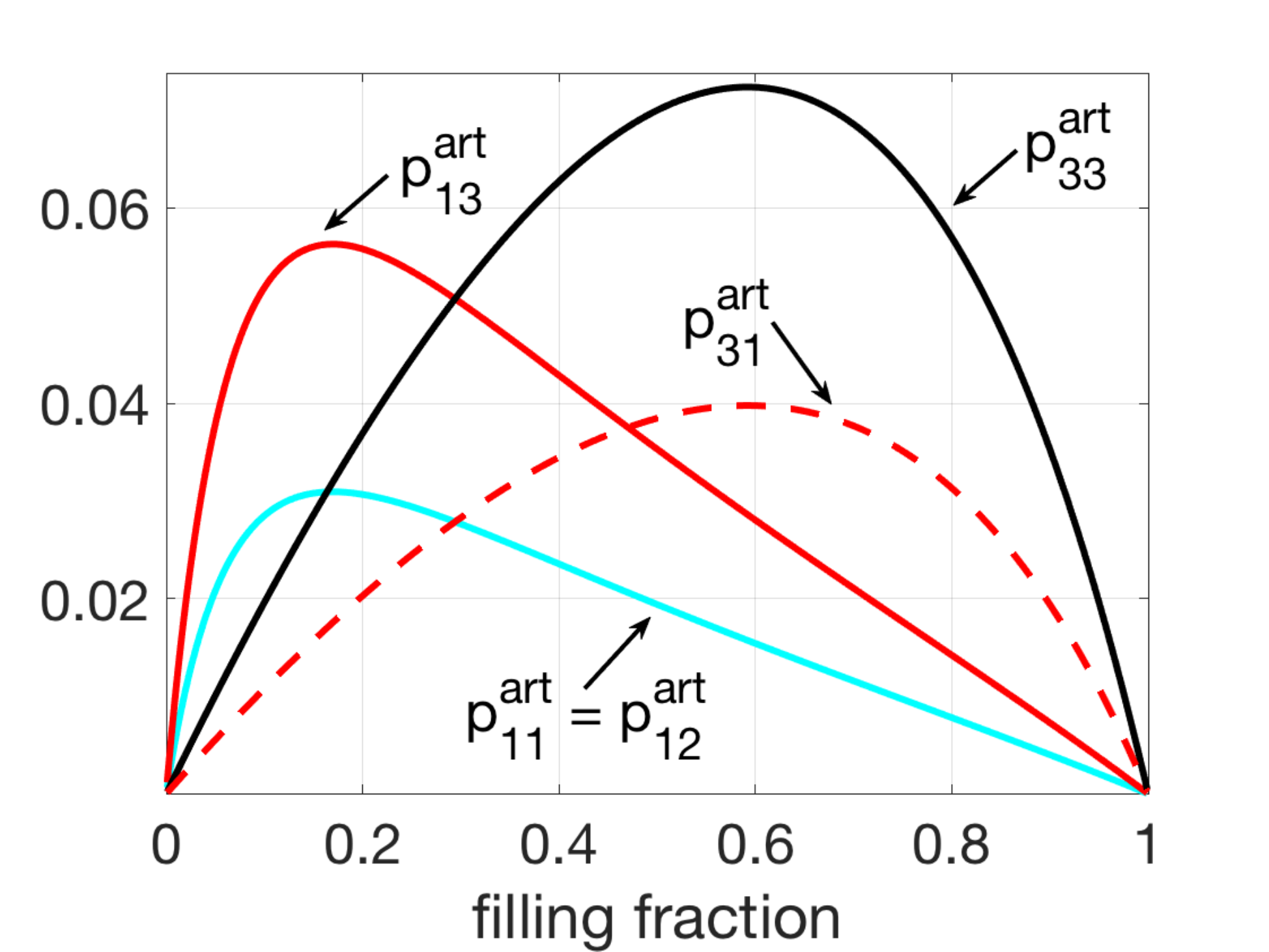}}
\subfigure[  \label{fig:sisio2C} ]{
\includegraphics[scale=0.28]{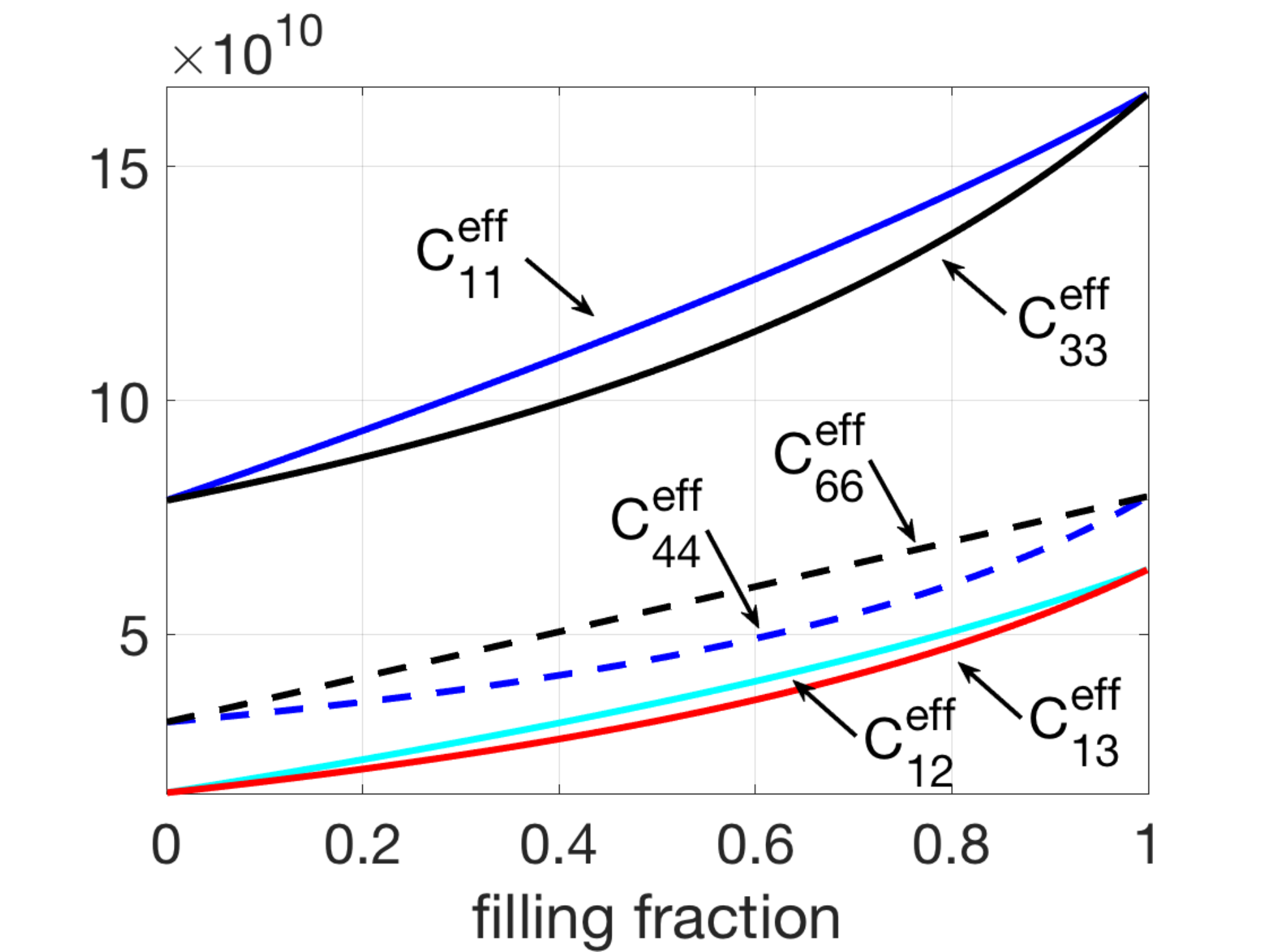}}
\subfigure[  \label{fig:sisio2vareps} ]{
\includegraphics[scale=0.28]{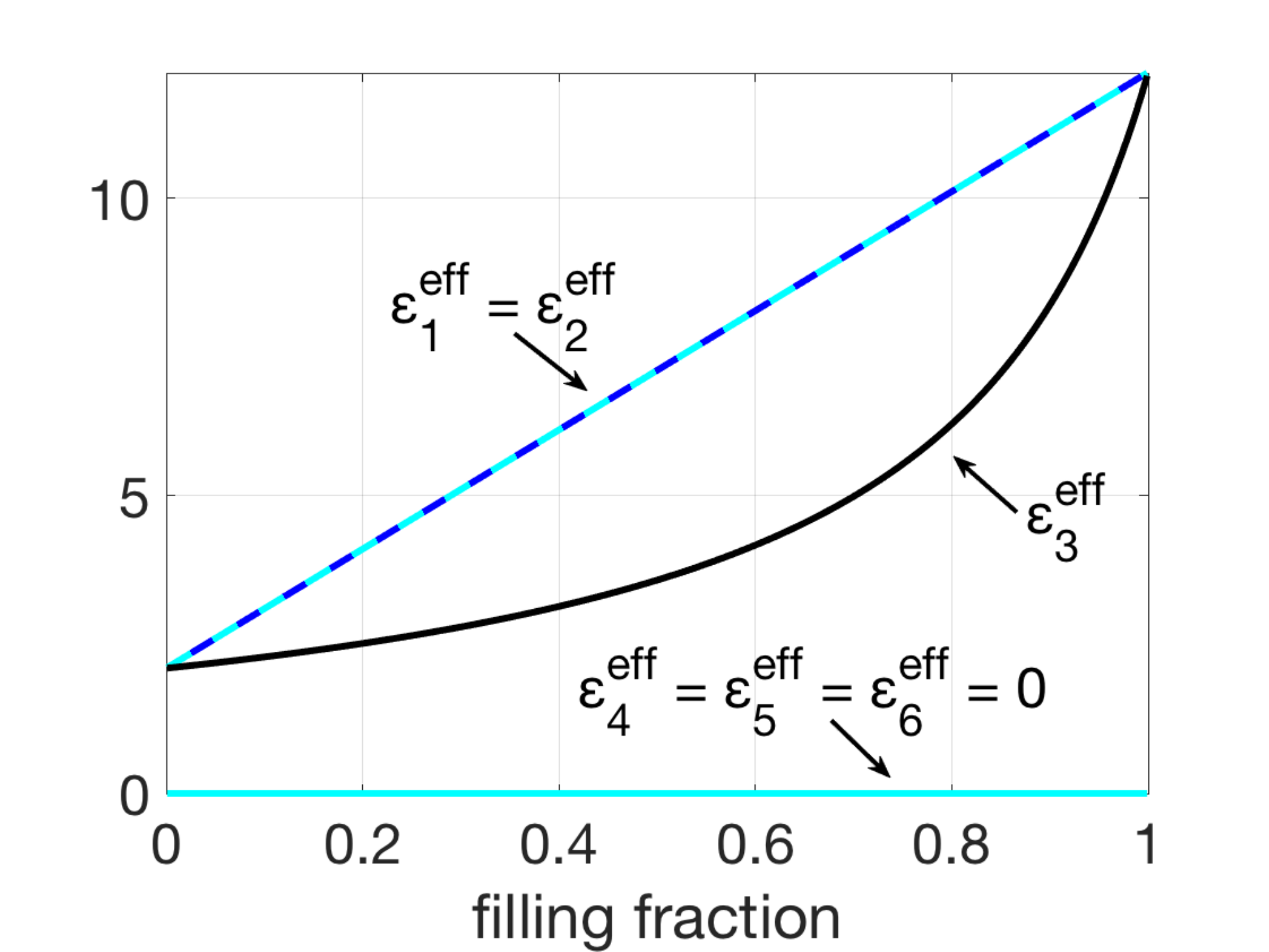}}
\subfigure[  \label{fig:sisio2gamma} ]{
\includegraphics[scale=0.28]{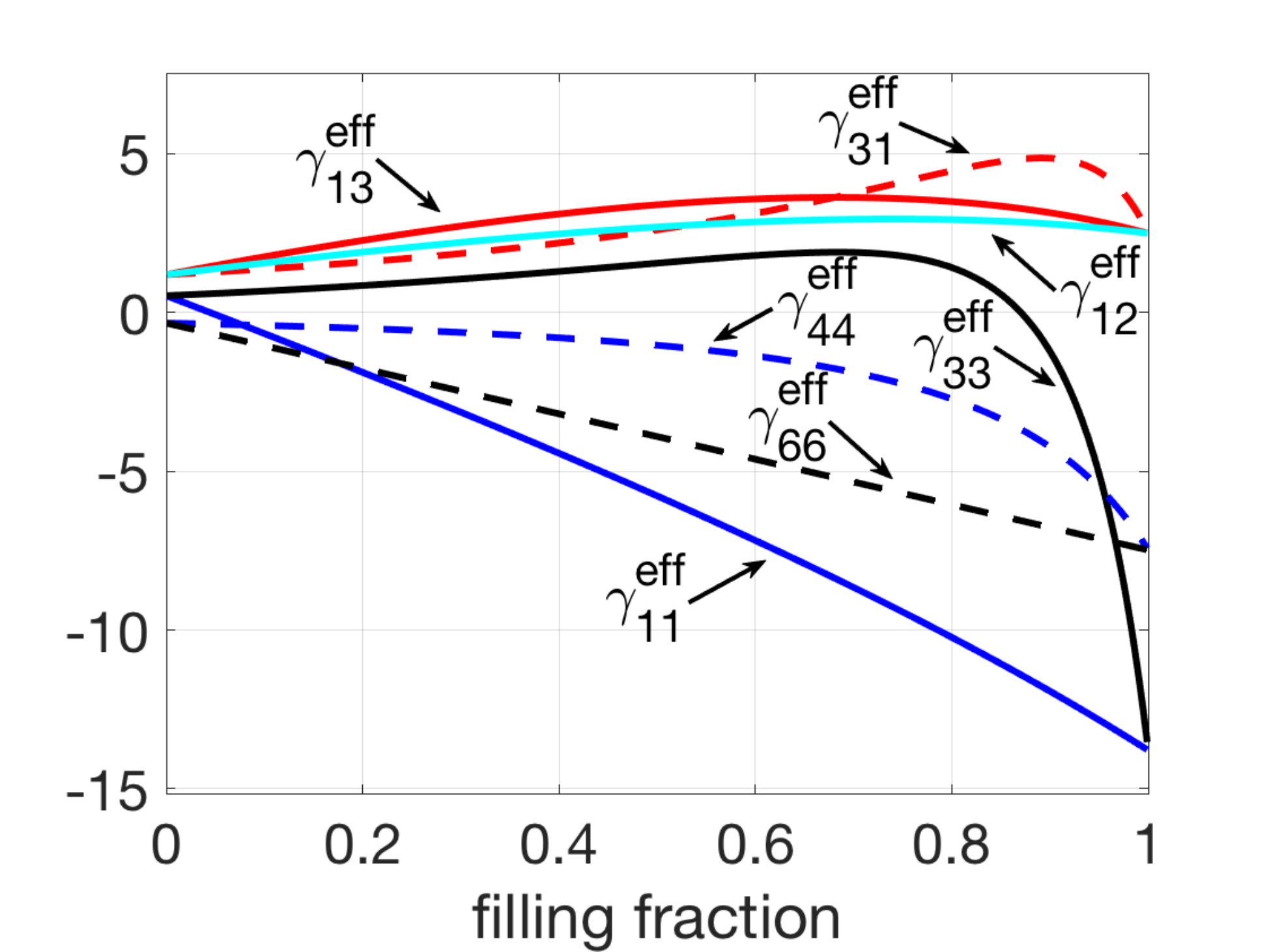}}

\caption{\label{fig:sisio2all}  \subref{fig:sisio2p1}   Symmetric photoelastic coefficients $p_{ijkl}^\eff$ corresponding to simple strains; 
\subref{fig:sisio2p2}  Roto-optic coefficient $r_{\yz\yz}^\eff$, and symmetric photoelastic coefficients $p_{\yz\yz}^\eff$  and $p_{\xy\xy}^\eff$;  
\subref{fig:sisio2art} Artificial photoelastic terms  $p_{ijkl}^\art$;
\subref{fig:sisio2C} Stiffness tensor coefficients $C_{ijkl}^\eff$; 
\subref{fig:sisio2vareps} Permittivity   coefficients $\varepsilon_{ij}^\eff$; and 
\subref{fig:sisio2gamma} Symmetric electrostriction   coefficients $\gamma_{ijkl}^\eff$;
 as a function of filling fraction $f$ for silica   and  Si $\left[100\right]$    layers with $a + a^\prime = 50\,\mathrm{nm}$, and labels in Voigt notation. 
  }
  \end{figure}
 \end{center}

 We begin by considering the material properties of a fused silica and silicon $[100]$ layered medium in Fig. \ref{fig:sisio2all}. In Fig. \ref{fig:sisio2p1} we present the symmetric photoelastic coefficients $p_{\xx\xx}^\eff$ (blue curve), $p_{\xx\yy}^\eff$ (cyan curve), $p_{\xx\zz}^\eff$ (red curve), $p_{\zz\zz}^\eff$ (black curve), and $p_{\zz\xx}^\eff$ (dashed red curve)  as a function of filling fraction. Here the coefficients exhibits  a varied dependence on filling fraction, with enhancement in the $p_{\xx\xx}^\eff$ and $p_{\zz\zz}^\eff$ elements beyond either of the constituent values to $p_{\zz\zz}^\eff = 0.135$ at $f=0.275$ and $p_{\xx\xx}^\eff = -0.121$ at $f=0.32$. For reference, we   describe  such behaviour as extraordinary enhancement (i.e.,when a composite material   possesses material values that are beyond the values for either of the constituents).    Interestingly,   the off-diagonal elements $p_{\xx\yy}^\eff$, $p_{\xx\zz}^\eff$, and $p_{\zz\xx}^\eff$ do not demonstrate extraordinary  enhancement for this material combination. We also have $p_{\xx\xx}^\eff = 0$ at $f=0.045$ along with $p_{\zz\zz}^\eff = 0$ at $f=0.87$, which implies that   longitudinal acoustic waves travelling along   $x$ at $f=0.045$, and  longitudinal acoustic waves travelling along $z$ at $f=0.87$, will not alter the optical properties of the medium. Reassuringly, symmetry-required degeneracies are recovered at $f=0$ and $f=1$ where the layered medium returns to a uniform material (i.e., $p_{\xx\xx} = p_{\zz\zz}$ and $p_{\xx\yy} = p_{\xx\zz} = p_{\zz\xx}$ in a cubic or isotropic medium).  In Fig. \ref{fig:sisio2p2} we show the remaining symmetric photoelastic coefficients $p_{\yz\yz}^\eff$ and $p_{\xy\xy}^\eff$ as functions of filling fraction, in addition to the roto-optic tensor coefficient $r_{\xy\xy}^\eff$. Here we observe a strong roto-optic effect in the layered material, which reaches a maximum of $r_{\xy\xy}^\eff = 0.081$ at $f = 0.295$. This value is greater than $|p_{\xy\xy}^\mathrm{Si}| = 0.051$ and  $|p_{\xy\xy}^\mathrm{SiO_2}| = 0.075$ and is also different in sign, which demonstrates that the roto-optic effect can significantly alter the predicted change in permittivity for acoustic shear waves,   and should not be omitted {\it a priori} without careful consideration. The two photoelastic shear constants also exhibit extraordinary enhancement, taking extremal values of $p_{\yz\yz}^\eff = -0.04$ at $f = 0.545$ and $p_{\xy\xy}^\eff = -0.109$ at $f = 0.117$, and also possess the correct degeneracy at $f=0$ and $f=1$ (i.e., $p_{\yz\yz}=p_{\xy\xy}$ in a cubic or isotropic medium).
 
  In Fig. \ref{fig:sisio2art} we present the artificial contributions to the photoelastic tensors shown in Fig. \ref{fig:sisio2p1}, which all exhibit  significant, positive-valued contributions to the effective symmetric photoelastic coefficients. This artificial contribution (obtained by substituting $p_{ijkl} = p_{ijkl}^\prime = 0$ in \eqref{eq:allPIJKLEff}) diminishes the extreme range of $p_{\xx\xx}^\eff$, and shows that the individual weightings for the constituent coefficients can take values $|\alpha_{qrst}|, |\alpha_{qrst}^\prime|>1$. This demonstrates that  extraordinary enhancement is possible without artificial photoelasticity.  In the case of a layered medium, we remark that $p_{\xx\xx}^\art = p_{\xx\yy}^\art$ since these terms  both arise from in-plane strains and are related to the same in-plane permittivity and stiffness contrast.  For reference, maximum values are listed as follows; $p_{\xx\xx}^\art = p_{\xx\yy}^\art  = 0.031$ at $f=0.17$, $p_{\zz\zz}^\art  = 0.072$ and $p_{\zz\xx}^\art =0.04$ at $f=0.592$, and $p_{\xx\zz}^\art =0.056$ at $f=0.17$.  In Figs. \ref{fig:sisio2C} and \ref{fig:sisio2vareps}, we present the effective stiffness and permittivity coefficients   as functions of filling fraction, following their explicit definitions in \eqref{eq:effeps} and \eqref{eq:allCeff}. Here, the simple dependences on $f$ are visible,   extraordinary   enhancements are not observed, and the required material symmetries are recovered at $f=0$ and $f=1$. In effect, for the permittivity and stiffness tensors, all in-plane terms are   given by volume averaging and   all out-of-plane terms are given by the inverse of volume-averaged reciprocal values. 
  
  In Figure \ref{fig:sisio2gamma} we give the corresponding symmetric electrostriction coefficients, defined as $\gamma_{ijkl} = \varepsilon_{ii}\varepsilon_{jj} p_{ijkl}$. In an analogous manner to the photoelastic coefficients, the electrostrictive coefficients of a composite material also exhibit a non-trivial dependence on $f$, in addition to extraordinary enhancement. Maxima  of $\gamma_{\xx\yy}^\eff = 2.96$ at $f = 0.74$,   $\gamma_{\xx\zz}^\eff = 3.626$ at $f = 0.68$, and $\gamma_{\zz\xx}^\eff = 4.87$ at $f = 0.89$  are observed,  demonstrating that the choice of polarisation and propagation direction can have important implications for  SBS experiments in layered media.

\begin{figure}[t]
\centering
\subfigure[ \label{fig:sio2as2s3p1}]{
\includegraphics[scale=0.28]{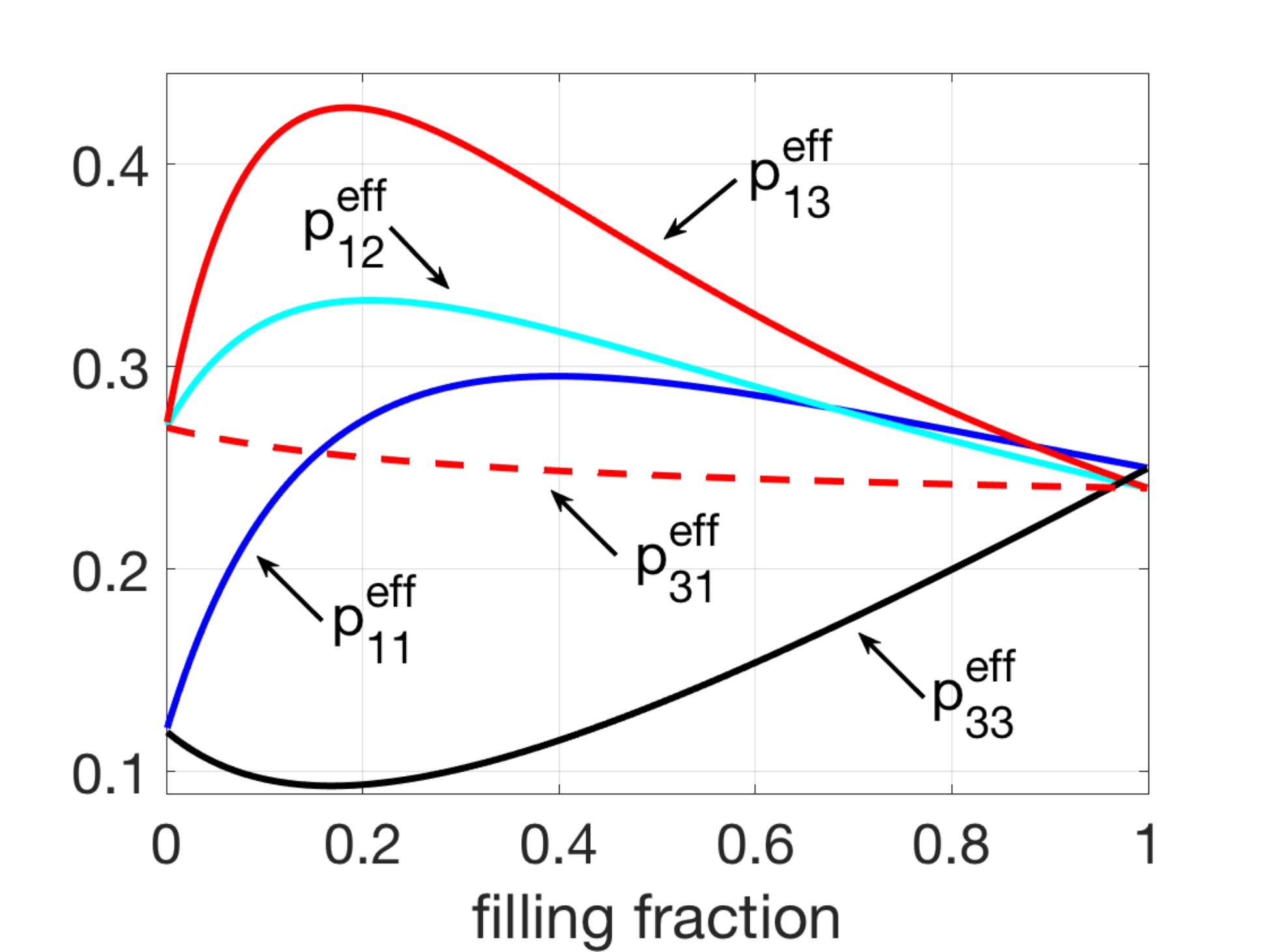}}
\subfigure[  \label{fig:sio2as2s3p2} ]{
\includegraphics[scale=0.28]{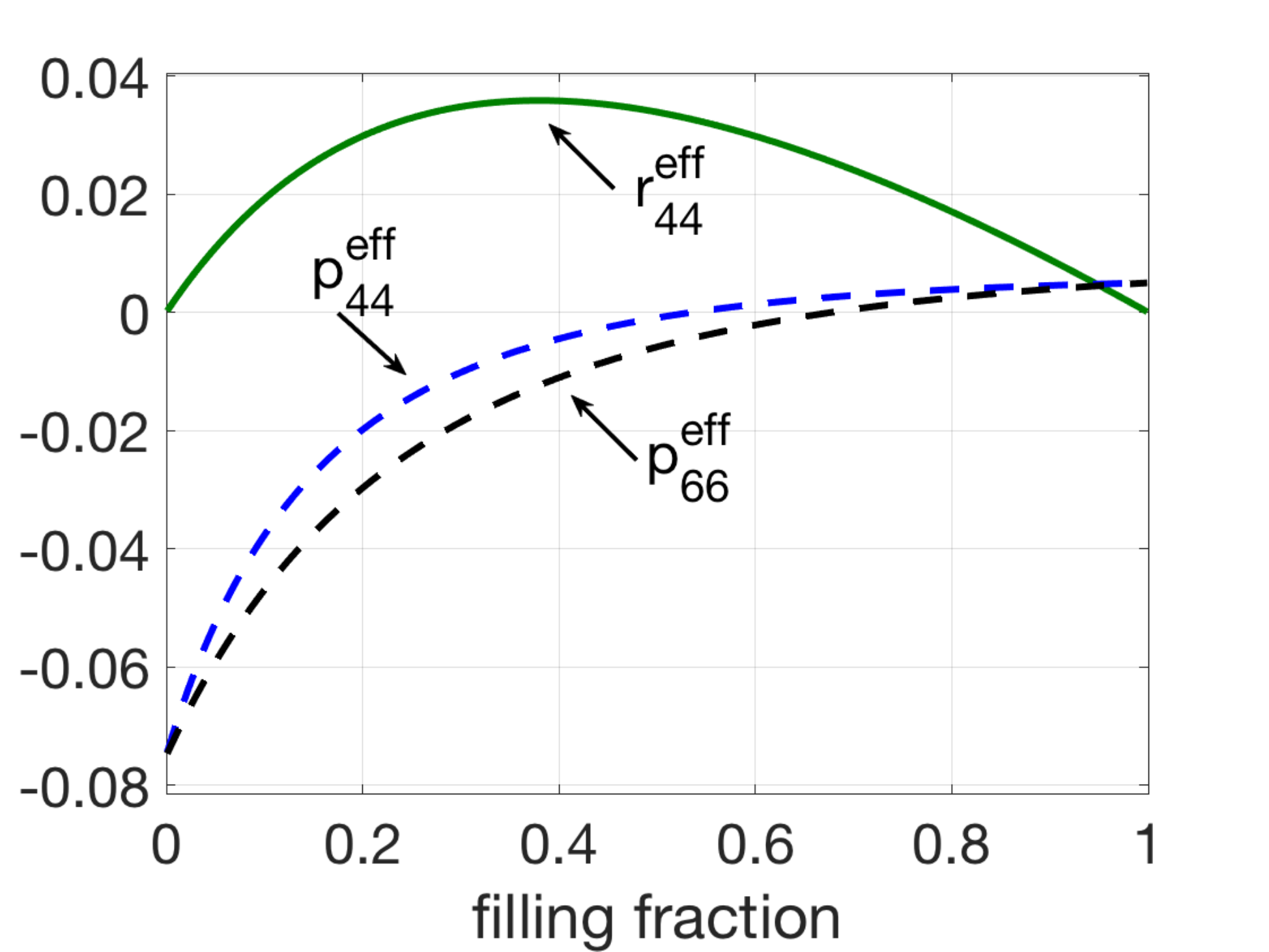}}
\subfigure[  \label{fig:sio2as2s3art} ]{
\includegraphics[scale=0.28]{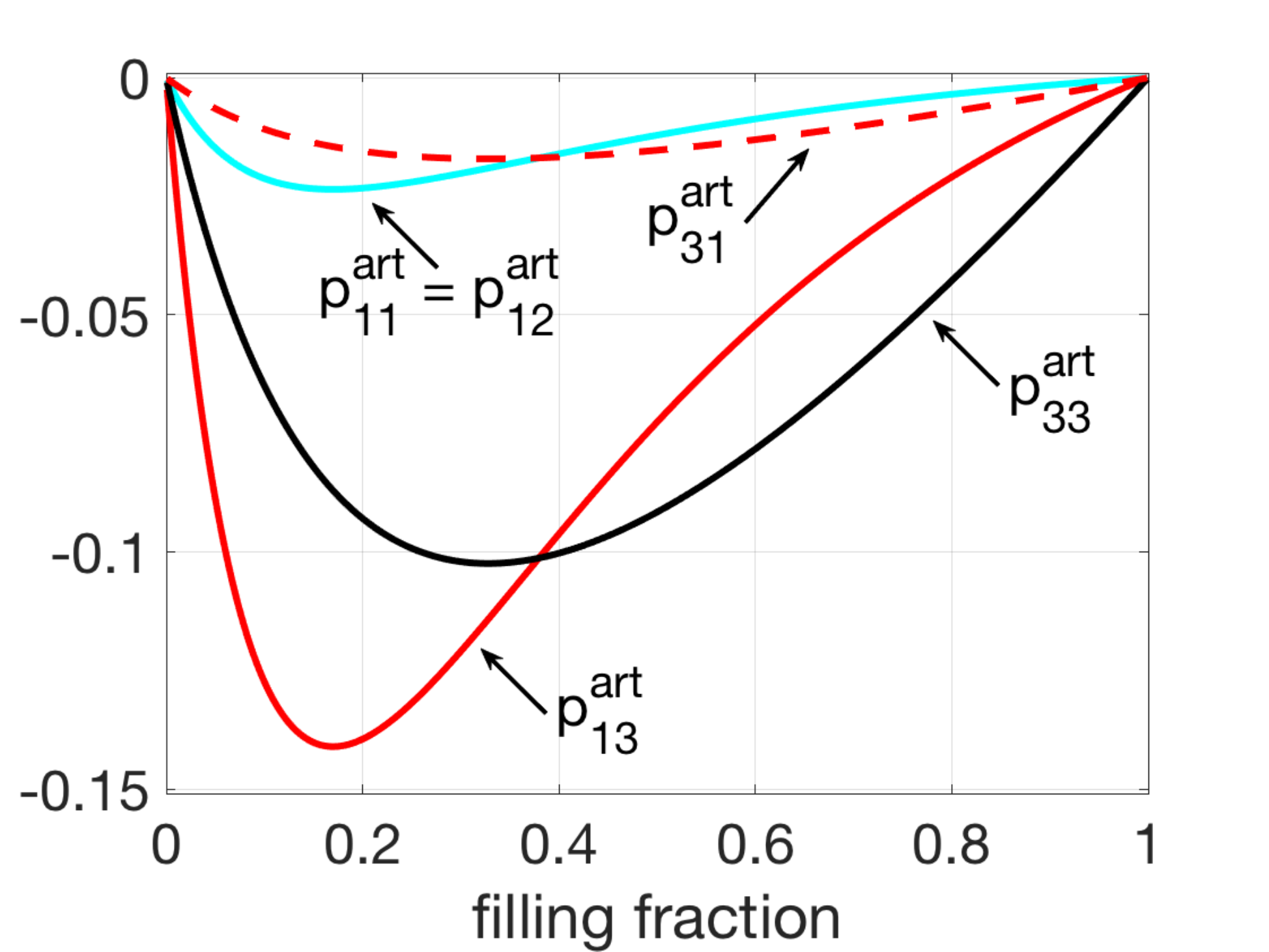}}
\subfigure[  \label{fig:sio2as2s3C} ]{
\includegraphics[scale=0.28]{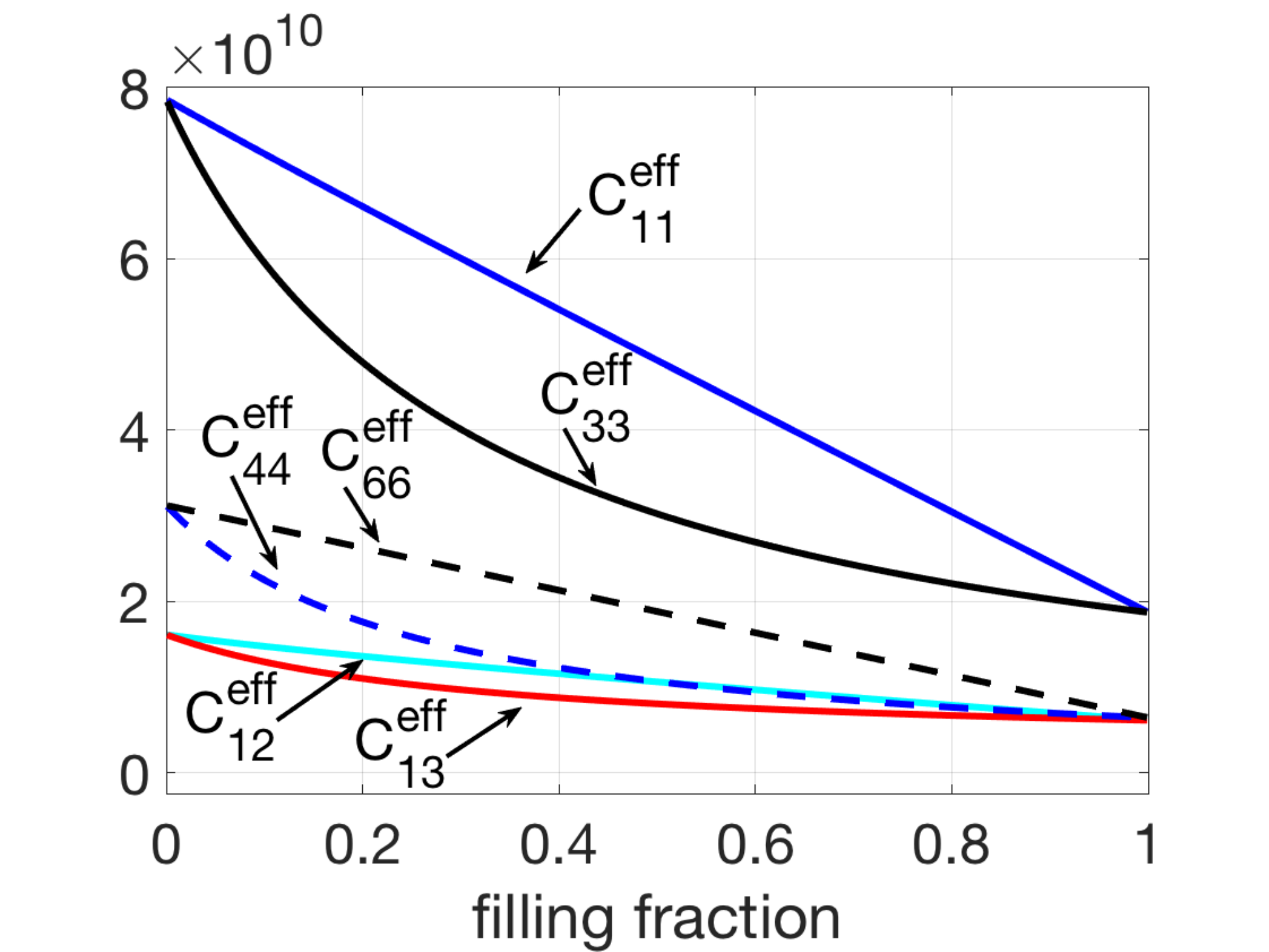}}
\subfigure[  \label{fig:sio2as2s3vareps} ]{
\includegraphics[scale=0.28]{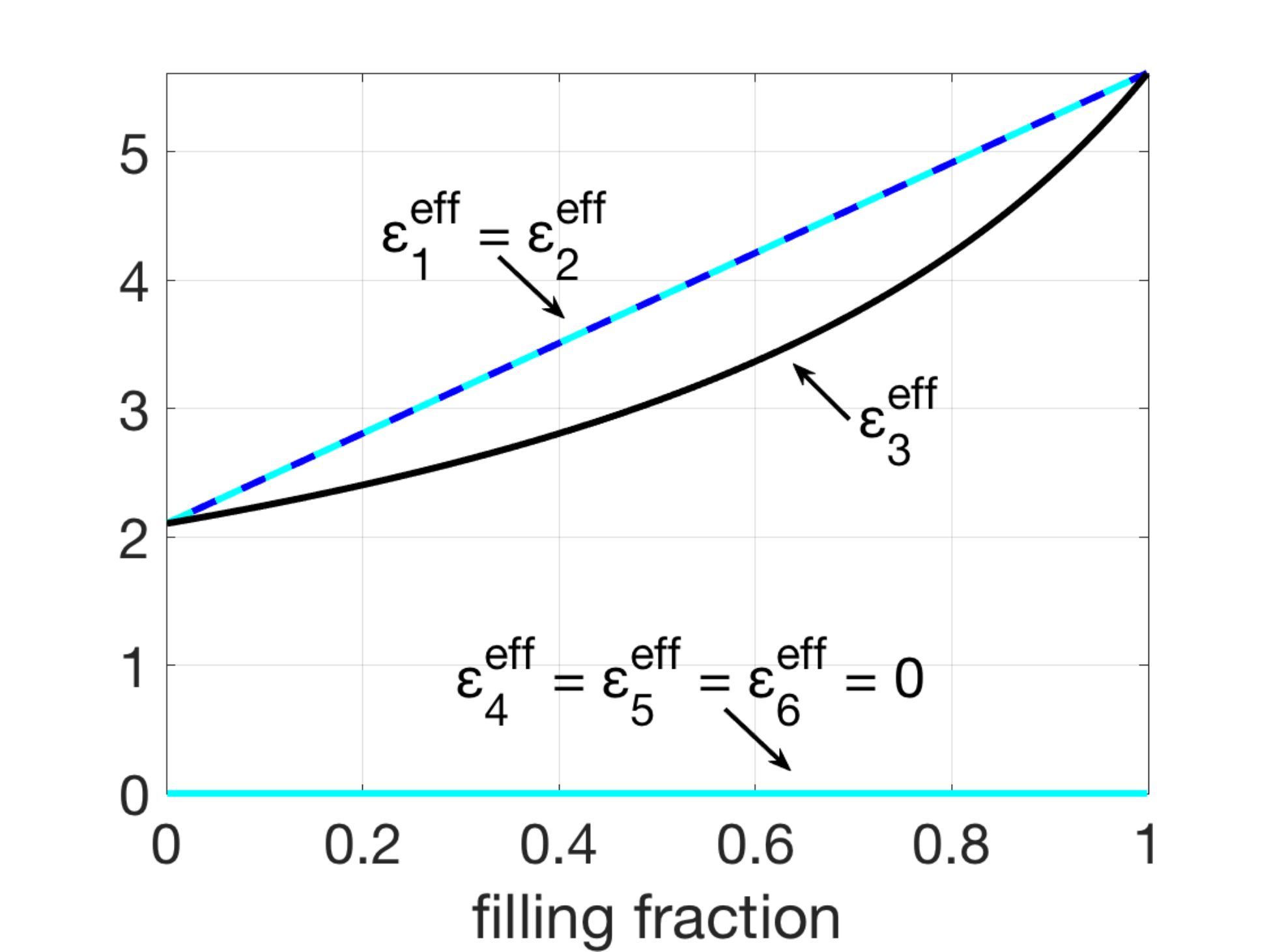}}
\subfigure[  \label{fig:sio2as2s3gamma} ]{
\includegraphics[scale=0.28]{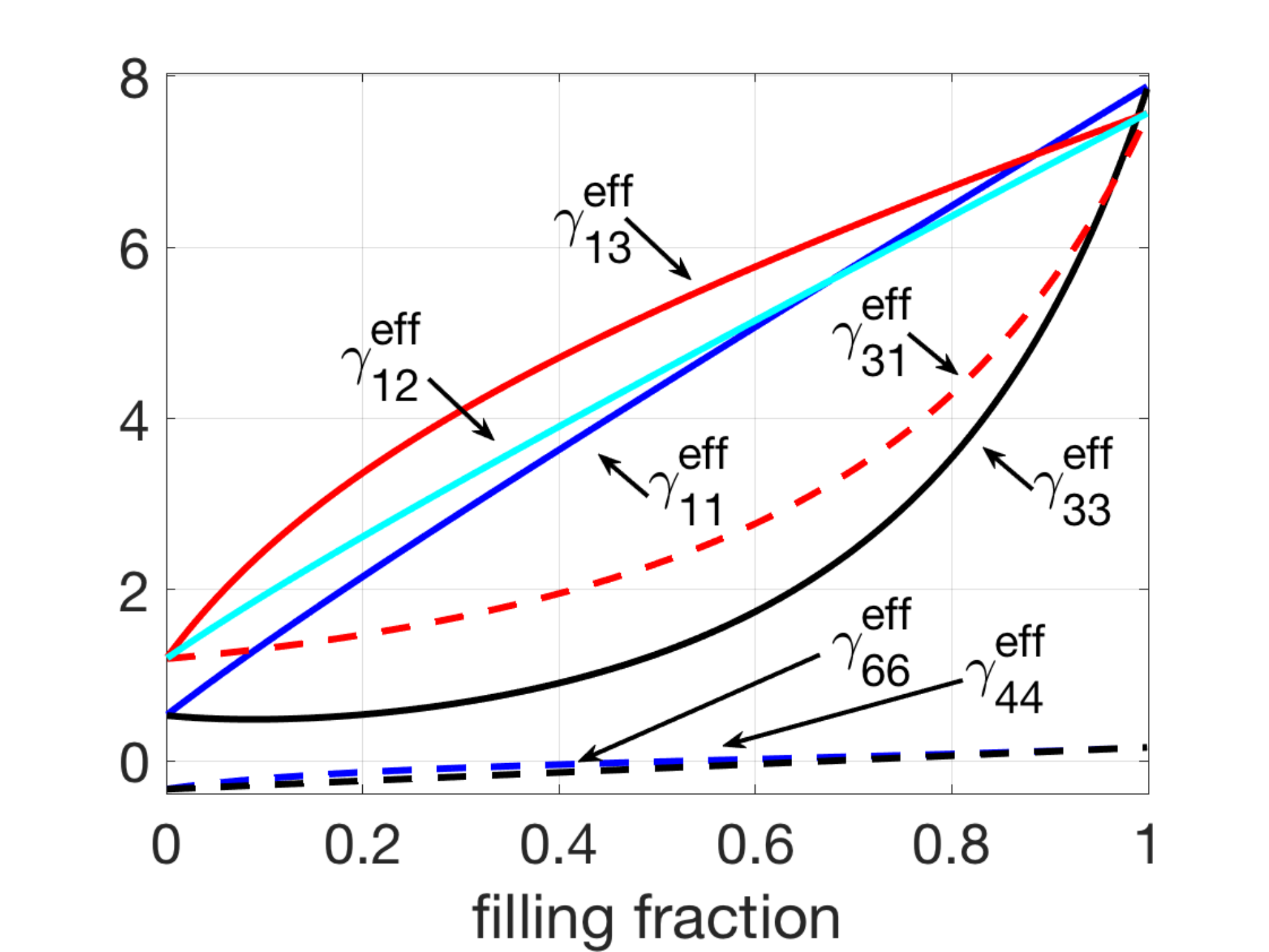}}

\caption{\label{fig:sio2as2s3all}  \subref{fig:sisio2p1}   Symmetric photoelastic coefficients $p_{ijkl}^\eff$ corresponding to simple strains; 
\subref{fig:sio2as2s3p2}  Roto-optic coefficient $r_{\yz\yz}^\eff$, and symmetric photoelastic coefficients $p_{\yz\yz}^\eff$  and $p_{\xy\xy}^\eff$;  
\subref{fig:sio2as2s3art} Artificial photoelastic terms  $p_{ijkl}^\art$;
\subref{fig:sio2as2s3C} Stiffness tensor coefficients $C_{ijkl}^\eff$; 
\subref{fig:sio2as2s3vareps} Permittivity   coefficients $\varepsilon_{ij}^\eff$; and 
\subref{fig:sio2as2s3gamma} Symmetric electrostriction   coefficients $\gamma_{ijkl}^\eff$;
 as a function of filling fraction $f$ for silica  and  As$_2$S$_3$-glass    layers with $a + a^\prime = 50\,\mathrm{nm}$, and labels in Voigt notation. 
  }
\end{figure}

In Fig. \ref{fig:sio2as2s3all} we present the material properties for a layered medium comprising fused silica and  As$_2$S$_3$-glass layers, in an analogous manner to Fig. \ref{fig:sisio2all}. In Fig. \ref{fig:sio2as2s3p1} we show a selection of symmetric photoelastic constants for the composite, where it is observed that all $p_{ijkl}^\eff$ corresponding to simple strains exhibit extraordinary enhancement. The enhancement of the $p_{\xx\zz}^\eff$ coefficient to  $p_{\xx\zz}^\eff = 0.428$ at $f=0.184$ is remarkable, when   compared to $p_{\xx\yy}^{\rmS\rmi\rmO_2} = 0.27$ and $p_{\xx\yy}^{\rmA\rms_2\rmS_3} = 0.24$ (i.e., an enhancement  of $59$\% and $78$\% respectively). In Fig. \ref{fig:sio2as2s3p2} we show the dependence on filling fraction for the remaining symmetric photoelastic constants, in addition to the roto-optic coefficient. For   silica and chalcogenide glass layers, a maximum of $r_{\yz\yz}^\eff = 0.036$ is achieved at $f=0.38$. Here we observe $p_{\yz\yz}^\eff = 0$ at $f = 0.535$ and $p_{\xy\xy}^\eff = 0$ at $f = 0.678$, implying that    acoustic shear waves   will travel through the material undetected when measuring the refractive index of the medium.

In Fig. \ref{fig:sio2as2s3art} we present the artificial contribution to the total symmetric photoelastic coefficients shown in Fig. \ref{fig:sio2as2s3p1}. Here it is apparent that artificial terms contribute negatively to the photoelastic properties of the layered medium, reducing the   $p_{\xx\zz}^\eff$ and $p_{\zz\zz}^\eff$ coefficients significantly. From this figure, we   also determine that the extraordinary enhancement in $p_{\xx\zz}^\eff$ is due to $|\alpha_{qrst}|,|\alpha_{qrst}^\prime|>1$. For reference, a maximum value  of $p_{\xx\zz}^\art = -0.141$ is achieved at $f=0.17$. In Fig. \ref{fig:sio2as2s3C} we present the stiffness tensor coefficients for the layered medium, and in Fig. \ref{fig:sio2as2s3vareps} we show the permittivity  tensor, as functions of filling fraction. Both of these figures exhibit a qualitatively similar behaviour to Figs. \ref{fig:sisio2C} and \ref{fig:sisio2vareps} with an absence of extraordinary enhancement. In Fig. \ref{fig:sio2as2s3gamma} we present the electrostriction constants as functions of filling fraction for this material combination, for completeness. Despite the large value for $p_{\xx\zz}^\eff$ observed in Fig. \ref{fig:sio2as2s3p1}, the corresponding $\gamma_{\xx\zz}^\eff$ term is smoothed by the much stronger growth in $(\varepsilon_\xx^\eff)^2$. Also, we observe zero values for  $\gamma_{\yz\yz}^\eff $ and $\gamma_{\xy\xy}^\eff $ following Fig. \ref{fig:sio2as2s3p2} along with   $\gamma_{\xx\xx}^\eff $ and  $\gamma_{\zz\zz}^\eff $ following Fig. \ref{fig:sio2as2s3p1}.

Following earlier works by some of the authors \cite{smith2016control,smith2016stimulated} on the numerical study of photoelasticity in composites comprising arrays of spheres, we now briefly compare results for a layered structure of silicon and chalcognide glass with   a corresponding cubic lattice structure. The  numerical procedure for the sphere configuration determines the effective bulk photoelastic response (including artificial contributions) by   comparing the change in the effective permittivity tensor relative to a small mechanical strain imposed on the unit cell boundary.

\begin{figure}[t]
\centering
\subfigure[ \label{fig:p11comp}]{
\includegraphics[scale=0.28]{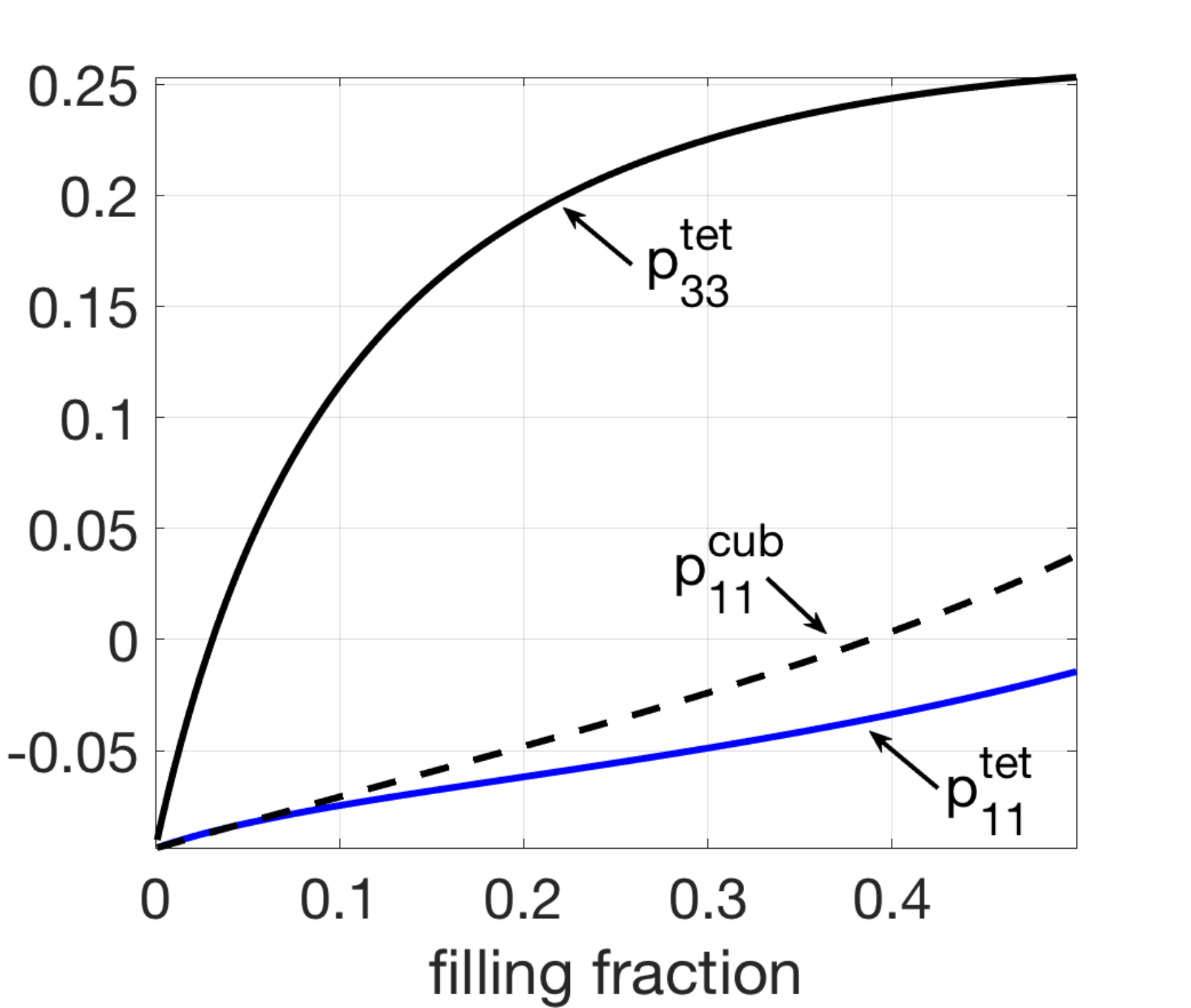}}
\subfigure[  \label{fig:p12comp} ]{
\includegraphics[scale=0.28]{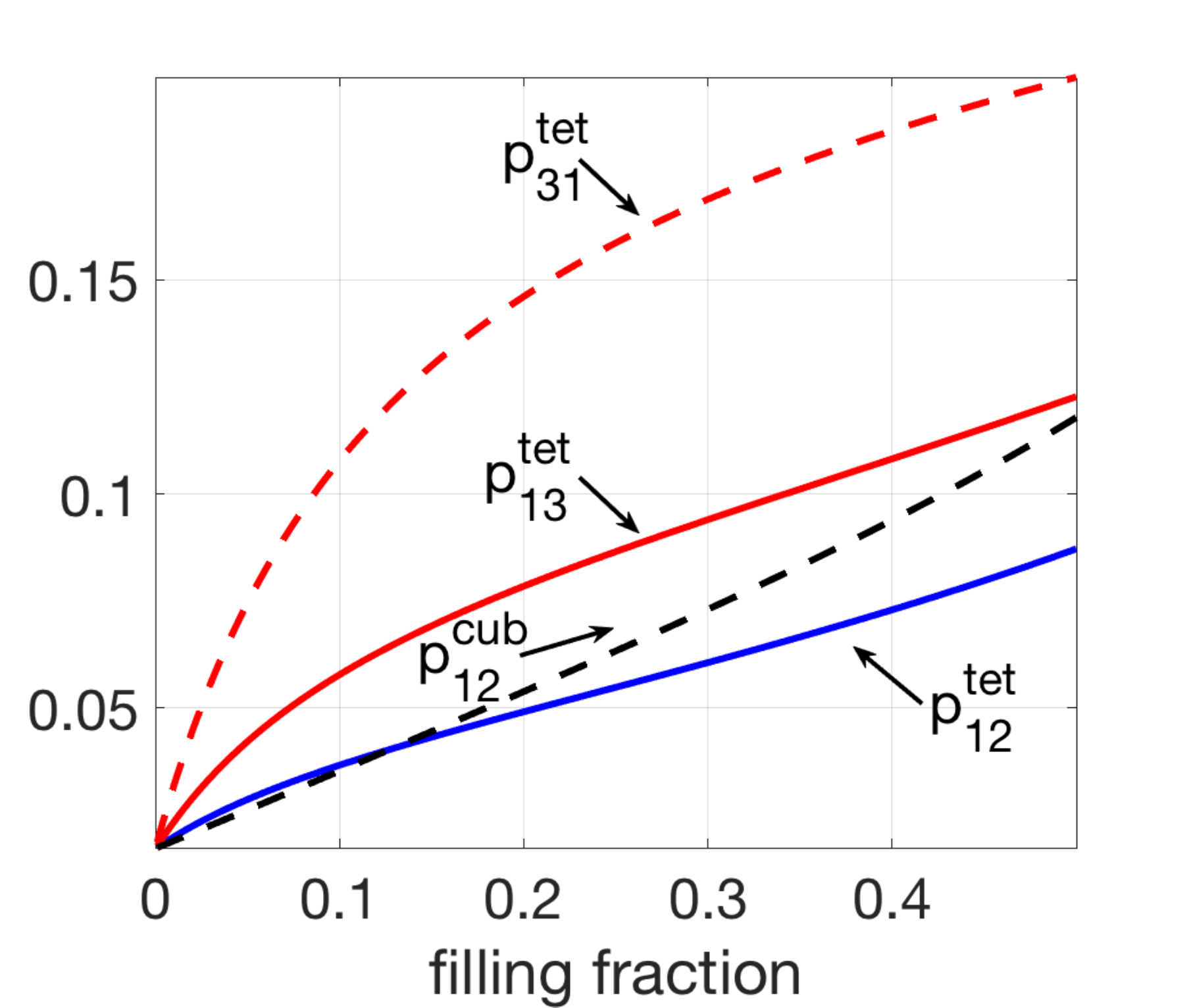}}
 
\caption{\label{fig:pcomp}  Comparison   of symmetric photoelastic coefficients  \subref{fig:p11comp}  $p_{\xx\xx}^\eff$; and  \subref{fig:p12comp} $p_{\xx\yy}^\eff$ as  function of filling fraction for  cubic array (cub) of  As$_2$S$_3$-glass spheres embedded in Si  $\left[100\right]$ (where $f = 4 \pi r^3/(3(a+a^\prime)^3)$,    $r$ is   radius of   sphere) with     corresponding terms for    layered medium (tet) comprising the same materials (with $f=a/(a+a^\prime)$),     $a+a^\prime=50\,\nm$. }
\end{figure}

 In Fig. \ref{fig:pcomp} we compare the photoelastic constants obtained with silicon and chalcogenide glass as a function of filling fraction, when these are structured in the form of a cubic array of spheres (cub) and as a layered material (tet). The values for the cubic material are obtained using an extended finite-element simulation procedure \cite{smith2016control,smith2016stimulated} where we restrict our attention to $0<f<0.5$ as this approaches the sphere touching limit and subsequently the extent of the numerical procedure.  In Figs. \ref{fig:p11comp} and \ref{fig:p12comp} we observe  that values for  the layered medium act as approximate bounds for the cubic lattice,  and suggest that our closed-form expressions may be used to obtain   estimates  of the photoelastic constants for an arbitrary material pair. The limit behaviour of these curves also   differs considerably with only $p_{\xx\xx}^\mathrm{tet}\approx p_{\xx\xx}^\mathrm{cub}$ and $p_{\xx\yy}^\mathrm{tet}\approx p_{\xx\yy}^\mathrm{cub}$ for vanishing filling fraction. However, we remark that further investigation is needed to determine bounds on the photoelastic properties of composite materials.

\onecolumngrid
  \clearpage

  \twocolumngrid

  \section{Concluding Remarks} \label{sec:conclrem} \noindent
 We have presented an accurate procedure for determining the acousto-optic properties of layered media, fully accounting for artificial photoelasticity and the roto-optic effect.   The methods   outlined in this work are fully consistent, transparent, and   easily generalisable to layered media with anisotropic constituents. This opens the path for exploring the acousto-optic properties of highly anisotropic media, such as hyperbolic metamaterials \cite{DraPodKil13} and thin film composites \cite{milton2002theory}.
  
  We   show that the  symmetric photoelastic constants $p_{ijkl}^\eff$ of a layered material are non-trival functions of filling fraction,   can exhibit extraordinary enhancement, and can be tuned as desired for applications. 
  
  We have also demonstrated that roto-optic coefficients can take comparable values to the symmetric photoelastic coefficients. This has important implications for acoustic shear wave propagation in optically anisotropic media. Furthermore, the tuneable photoelastic response offered by layered materials may    have important  implications for   SBS structures.

 \section*{Acknowledgements }
This work was supported by the Australian Research Council: CUDOS Centre of Excellence CE110001018, and Discovery Projects DP150103611, DP160101691. 

 \bibliography{layered_pe}

\end{document}